\pdfoutput=1
\documentclass[12pt,a4paper]{article}

\usepackage{microtype}
\usepackage{lineno}  
\usepackage{xspace} 

\usepackage{graphicx}  
\usepackage{color}
\usepackage{colortbl}

\usepackage{amsmath} 
\usepackage{ifthen} 
\usepackage{subfigure}
\usepackage{bm}
\usepackage{multirow}

\newboolean{pdflatex}
\setboolean{pdflatex}{true} 

\newboolean{articletitles}
\setboolean{articletitles}{true} 

\newboolean{uprightparticles}
\setboolean{uprightparticles}{false} 

\usepackage{amssymb}
\usepackage{amsfonts}
\usepackage{upgreek} 

\newcommand*\patchAmsMathEnvironmentForLineno[1]{%
\expandafter\let\csname old#1\expandafter\endcsname\csname #1\endcsname
\expandafter\let\csname oldend#1\expandafter\endcsname\csname
end#1\endcsname
 \renewenvironment{#1}%
   {\linenomath\csname old#1\endcsname}%
   {\csname oldend#1\endcsname\endlinenomath}%
}
\newcommand*\patchBothAmsMathEnvironmentsForLineno[1]{%
  \patchAmsMathEnvironmentForLineno{#1}%
  \patchAmsMathEnvironmentForLineno{#1*}%
}
\AtBeginDocument{%
\patchBothAmsMathEnvironmentsForLineno{equation}%
\patchBothAmsMathEnvironmentsForLineno{align}%
\patchBothAmsMathEnvironmentsForLineno{flalign}%
\patchBothAmsMathEnvironmentsForLineno{alignat}%
\patchBothAmsMathEnvironmentsForLineno{gather}%
\patchBothAmsMathEnvironmentsForLineno{multline}%
}

	\usepackage{hyperref}    
\usepackage[all]{hypcap} 

\def\Kstarzz  {\ensuremath{\kaon^{*0}_{\mathrm{0}}}\xspace}
\def\Kstarzo  {\ensuremath{\kaon^{*0}_{\mathrm{1}}}\xspace}
\def\Kstarzt  {\ensuremath{\kaon^{*0}_{\mathrm{2}}}\xspace}

\def\FL    {\ensuremath{F_{\mathrm{L}}}\xspace}
\def\FS    {\ensuremath{F_{\mathrm{S}}}\xspace}

\def\AS    {\ensuremath{A_{\mathrm{S}}}\xspace}
\def\AIm   {\ensuremath{A_{\mathrm{Im}}}\xspace}

\def\kpi                {\ensuremath{{K^{+}\pi^{-}}}\xspace}
\def\BdToKstll    {\decay{\Bd}{\Kstarz\ellp\ellm}}

\def\BdToKpill    {\decay{\Bd}{\Kp\pim\ellp\ellm}}
\def\phiprime {\ensuremath{\phi^{'}}\xspace}
\def\FS    {\ensuremath{F_{\mathrm{S}}}\xspace}
\def\FSi    {\ensuremath{\mathcal{F}_{\mathrm{S}}}\xspace}
\def\FP    {\ensuremath{F_{\mathrm{P}}}\xspace}
\def\FPi   {\ensuremath{\mathcal{F}_{\mathrm{P}}}\xspace}

\def\AS    {\ensuremath{A_{\mathrm{S}}}\xspace}
\def\ASi    {\ensuremath{\mathcal{A}_{\mathrm{S}}}\xspace}

\def\psq    {\ensuremath{p^2}\xspace}

\def\lhcb {LHCb\xspace}
\def\ux85 {UX85\xspace}
\def\cern {CERN\xspace}
\def\lhc {LHC\xspace}

\def\babar  {BaBar\xspace}
\def\belle  {Belle\xspace}

\def\cdf    {CDF\xspace}

\ifthenelse{\boolean{uprightparticles}}%
{
 
 \def\Pgamma      {\ensuremath{\upgamma}\xspace}

 \def\Pmu         {\ensuremath{\upmu}\xspace}

 \def\Ppi         {\ensuremath{\uppi}\xspace}

 \def\Ppsi        {\ensuremath{\uppsi}\xspace}

 \def\PDelta      {\ensuremath{\Delta}\xspace}                 
 \def\PXi      {\ensuremath{\Xi}\xspace}                 
 \def\PLambda      {\ensuremath{\Lambda}\xspace}                 
 \def\PSigma      {\ensuremath{\Sigma}\xspace}                 
 \def\POmega      {\ensuremath{\Omega}\xspace}                 
 \def\PUpsilon      {\ensuremath{\Upsilon}\xspace}                 
 
 \def\PB      {\ensuremath{\mathrm{B}}\xspace}                 
                  
 \def\PD      {\ensuremath{\mathrm{D}}\xspace}

 \def\PJ      {\ensuremath{\mathrm{J}}\xspace}                 
 \def\PK      {\ensuremath{\mathrm{K}}\xspace}

 \def\Pb      {\ensuremath{\mathrm{b}}\xspace}

 \def\Pi      {\ensuremath{\mathrm{i}}\xspace}

 \def\Ps      {\ensuremath{\mathrm{s}}\xspace}

}
{
 
 \def\Pgamma      {\ensuremath{\gamma}\xspace}

 \def\Pmu         {\ensuremath{\mu}\xspace}

 \def\Ppi         {\ensuremath{\pi}\xspace}

 \def\Ppsi        {\ensuremath{\psi}\xspace}                 
                  
 \mathchardef\PDelta="7101
 \mathchardef\PXi="7104
 \mathchardef\PLambda="7103
 \mathchardef\PSigma="7106
 \mathchardef\POmega="710A
 \mathchardef\PUpsilon="7107
                  
 \def\PB      {\ensuremath{B}\xspace}                 
                  
 \def\PD      {\ensuremath{D}\xspace}

 \def\PJ      {\ensuremath{J}\xspace}                 
 \def\PK      {\ensuremath{K}\xspace}

 \def\Pb      {\ensuremath{b}\xspace}

 \def\Pi      {\ensuremath{i}\xspace}

 \def\Ps      {\ensuremath{s}\xspace}

}

\def\mup        {\ensuremath{\Pmu^+}\xspace}
\def\mun        {\ensuremath{\Pmu^-}\xspace} 

\def\ellm       {\ensuremath{\ell^-}\xspace}
\def\ellp       {\ensuremath{\ell^+}\xspace}
\def\ellell     {\ensuremath{\ell^+ \ell^-}\xspace}

\def\g      {\ensuremath{\Pgamma}\xspace}

\def\squark    {\ensuremath{\Ps}\xspace}

\def\bquark    {\ensuremath{\Pb}\xspace}

\def\pion  {\ensuremath{\Ppi}\xspace}

\def\pim   {\ensuremath{\pion^-}\xspace}

\def\kaon  {\ensuremath{\PK}\xspace}
  \def\Kbar  {\kern 0.2em\overline{\kern -0.2em \PK}{}\xspace}

\def\Kz    {\ensuremath{\kaon^0}\xspace}
\def\Kzb   {\ensuremath{\Kbar^0}\xspace}
\def\KzKzb {\ensuremath{\Kz \kern -0.16em \Kzb}\xspace}
\def\Kp    {\ensuremath{\kaon^+}\xspace}
\def\Km    {\ensuremath{\kaon^-}\xspace}

\def\KpKm  {\ensuremath{\Kp \kern -0.16em \Km}\xspace}

\def\Kstarz  {\ensuremath{\kaon^{*0}}\xspace}

\def\Kstar   {\ensuremath{\kaon^*}\xspace}

  \def\Dbar    {\kern 0.2em\overline{\kern -0.2em \PD}{}\xspace}
\def\D       {\ensuremath{\PD}\xspace}

\def\Dz      {\ensuremath{\D^0}\xspace}
\def\Dzb     {\ensuremath{\Dbar^0}\xspace}
\def\DzDzb   {\ensuremath{\Dz {\kern -0.16em \Dzb}}\xspace}
\def\Dp      {\ensuremath{\D^+}\xspace}
\def\Dm      {\ensuremath{\D^-}\xspace}

\def\DpDm    {\ensuremath{\Dp {\kern -0.16em \Dm}}\xspace}

\def\B       {\ensuremath{\PB}\xspace}
  \def\Bbar    {\kern 0.18em\overline{\kern -0.18em \PB}{}\xspace}

\def\Bz      {\ensuremath{\B^0}\xspace}

\def\Bd      {\ensuremath{\B^0}\xspace}
\def\Bs      {\ensuremath{\B^0_\squark}\xspace}

\def\jpsi     {\ensuremath{{\PJ\mskip -3mu/\mskip -2mu\Ppsi\mskip 2mu}}\xspace}

  \def\Y#1S{\ensuremath{\PUpsilon{(#1S)}}\xspace}

\newcommand{\decay}[2]{\ensuremath{#1\!\to #2}\xspace}         

\def\to                 {\ensuremath{\rightarrow}\xspace}

\def\qsq       {\ensuremath{q^2}\xspace}

\def\CP                {\ensuremath{C\!P}\xspace}

\def\BdToKstmm    {\decay{\Bd}{\Kstarz\mup\mun}}

\def\AFB      {\ensuremath{A_{\mathrm{FB}}}\xspace}
\def\FL       {\ensuremath{F_{\mathrm{L}}}\xspace}
\def\AT#1     {\ensuremath{A_{\mathrm{T}}^{#1}}\xspace}           
\def\btosgam  {\decay{\bquark}{\squark \g}}

\def\ctl       {\ensuremath{\cos{\theta_l}}\xspace}
\def\ctk       {\ensuremath{\cos{\theta_K}}\xspace}
\def\ctlsq    {\ensuremath{\cos^2{\theta_l}}\xspace}
\def\ctksq    {\ensuremath{\cos^2{\theta_K}}\xspace}
\def\stl      {\ensuremath{\sin{\theta_l}}\xspace}
\def\stk      {\ensuremath{\sin{\theta_K}}\xspace}
\def\stlsq    {\ensuremath{\sin^2{\theta_l}}\xspace}
\def\stksq    {\ensuremath{\sin^2{\theta_K}}\xspace}
\def\dctl       {\ensuremath{\mathrm{dcos}{\theta_l}}\xspace}
\def\dctk       {\ensuremath{\mathrm{dcos}{\theta_K}}\xspace}

\def\C#1      {\ensuremath{\mathcal{C}_{#1}}\xspace}                       
\def\Cp#1     {\ensuremath{\mathcal{C}_{#1}^{'}}\xspace}                    
\def\Ceff#1   {\ensuremath{\mathcal{C}_{#1}^{\mathrm{(eff)}}}\xspace}        
\def\Cpeff#1  {\ensuremath{\mathcal{C}_{#1}^{'\mathrm{(eff)}}}\xspace}       
\def\Ope#1    {\ensuremath{\mathcal{O}_{#1}}\xspace}                       
\def\Opep#1   {\ensuremath{\mathcal{O}_{#1}^{'}}\xspace}                    

\def\kpi        {\ensuremath{\Kp\pim}\xspace}

\newcommand{\tev}{\ensuremath{\mathrm{\,Te\kern -0.1em V}}\xspace}
\newcommand{\gev}{\ensuremath{\mathrm{\,Ge\kern -0.1em V}}\xspace}
\newcommand{\mev}{\ensuremath{\mathrm{\,Me\kern -0.1em V}}\xspace}
\newcommand{\kev}{\ensuremath{\mathrm{\,ke\kern -0.1em V}}\xspace}
\newcommand{\ev}{\ensuremath{\mathrm{\,e\kern -0.1em V}}\xspace}
\newcommand{\gevc}{\ensuremath{{\mathrm{\,Ge\kern -0.1em V\!/}c}}\xspace}
\newcommand{\mevc}{\ensuremath{{\mathrm{\,Me\kern -0.1em V\!/}c}}\xspace}
\newcommand{\gevcc}{\ensuremath{{\mathrm{\,Ge\kern -0.1em V\!/}c^2}}\xspace}
\newcommand{\gevgevcccc}{\ensuremath{{\mathrm{\,Ge\kern -0.1em V^2\!/}c^4}}\xspace}
\newcommand{\mevcc}{\ensuremath{{\mathrm{\,Me\kern -0.1em V\!/}c^2}}\xspace}
\newcommand{\mevmevcccc}{\ensuremath{{\mathrm{\,Me\kern -0.1em V^2\!/}c^4}}\xspace}

\def\deriv {\ensuremath{\mathrm{d}}}

\def\gsim{{~\raise.15em\hbox{$>$}\kern-.85em
          \lower.35em\hbox{$\sim$}~}\xspace}
\def\lsim{{~\raise.15em\hbox{$<$}\kern-.85em
          \lower.35em\hbox{$\sim$}~}\xspace}

\def\tell1  {TELL1\xspace}
\def\ukl1   {UKL1\xspace}

\makeatletter
\DeclareRobustCommand*{\bfseries}{%
  \not@math@alphabet\bfseries\mathbf
  \fontseries\bfdefault\selectfont
  \boldmath
}
\makeatother

\numberwithin{equation}{section}

\usepackage{cite}
\usepackage{mciteplus}

\begin{document}

\begin{titlepage}
\pagenumbering{roman}

\hspace*{-0.5cm}
\begin{tabular*}{\linewidth}{lc@{\extracolsep{\fill}}r}
\vspace*{-1.0cm}\mbox{\!\!\!\includegraphics[width=.20\textwidth]{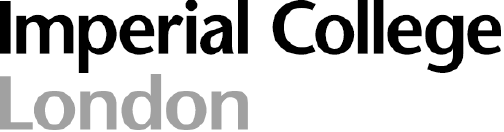}} & & %
\\
 & & IC/HEP/12-03 \\  
 & & March 18, 2013 \\ 
 & & \\
\end{tabular*}

\vspace*{2.0cm}

{\bf\boldmath\huge
\begin{center}
The effect of S-wave interference on the \BdToKstll angular observables
\end{center}
}

\vspace*{2.0cm}

{
\begin{center}
\today
\end{center}
}

\begin{center}
  Thomas Blake$^a$, Ulrik Egede$^b$, Alex Shires$^b$\footnote{Corresponding author.}\\
  \emph{$^a$ CERN, Geneva, Switzerland} \\
  \emph{$^b$ Imperial College London, London SW7 2AZ, United Kingdom}
\end{center}

\begin{abstract}
  \noindent
  The rare decay \BdToKstll is a flavour changing neutral current
  decay with a high sensitivity to physics beyond the Standard Model.
  Nearly all theoretical predictions and all experimental measurements 
  so far have assumed a \Kstarz P-wave that decays into the \kpi final 
  state. In this paper the addition of an S-wave within the \kpi system 
  of \BdToKstll and the subsequent impact of this on the angular 
  distribution of the final state particles is explored.  
  The inclusion of the S-wave causes a distinction between 
  the values of the angular observables obtained
  from counting experiments and those obtained from fits to the
  angular distribution. The effect of a non-zero S-wave on an angular
  analysis of \BdToKstll is assessed as a function of dataset size and
  the relative size of the S-wave amplitude.  An S-wave
  contribution,
  equivalent to what is measured in $\Bd\to\jpsi\Kstarz$ at \babar,
  leads to a significant bias on the angular
  observables for datasets of above  200 signal decays. 
  Any future experimental analysis of the 
  $\kpi\ellell$ final state will have to take the S-wave contribution
   into account.
\end{abstract}

\vspace*{1.0cm}
\begin{center}
\end{center}
\vspace{\fill}

\end{titlepage}

\pagestyle{plain} 
\setcounter{page}{1}
\pagenumbering{arabic}

\newpage

\section{Introduction}
\label{sec:Introduction}

The description of flavour physics in the Standard Model (SM) has so 
far accurately matched the observations in the data from the \B factories, 
the Tevatron and the \lhc very well. 
However, there are several fundamental questions which do not have 
an explanation within the SM such as the
mass hierarchy of the quarks and why there are three generations.
To avoid creating large flavour changing neutral currents, any physics 
beyond the SM that contains new degrees of freedom that couple to the flavour sector
is required to be at an
energy scale of multiple \tev or to have small couplings between the generations, 
i.e. couplings that closely mimic those of the SM. 
The measurement of the inclusive
\btosgam width~\cite{Asner:2010qj} is one of the strongest constraints 
on new physics from the flavour sector; 
for the exclusive decays, \BdToKstll is of major importance. 

The analysis of \BdToKstll is based on the evaluating the angular
distribution of the daughter particles~\cite{PhysRevD.61.114028}. 
How to extract the maximal amount of information from the decay
while keeping uncertainties from QCD minimal has recently attracted much 
interest~\cite{Kruger:2005ep,Egede:2008uy,AltmannshoferBall,Egede:2010zc,Bobeth:2010wg,Matias:2012xw}.
The results from the experimental analyses of 
\BdToKstll~\cite{Aubert:2008ju,PhysRevLett.103.171801,Aaltonen:2011cn,Aaij:2011aa} 
have focused on the forward backward asymmetry of the
dimuon system (\AFB) and the fraction of longitudinal polarisation of
the \Kstarz (\FL) as a function of the dimuon invariant mass.

With the acquisition of large data sets of \BdToKstll decays,
scrutiny is required of assumptions that have been made in current experiments. 
Nearly all theoretical papers to date use the narrow width
assumption for the \kpi system meaning that the natural width of
the $\Kstarz(892)$ is ignored. 
This means there is no interference with other
\kpi resonances. Existing \BdToKstll analyses
consider \BdToKstll signal with \kpi candidates in a
narrow mass window around the $\Kstarz(892)$. However, in this region
there is evidence of a broad S-wave below the $\Kstarz(892)$
and higher mass states which decay strongly to \kpi, such as the S-wave $\Kstarzz(1430)$
and the D-wave $\Kstarzt(1430)$~\cite{Beringer:1900zz}. 
The best understanding of the low mass S-wave contribution comes from
the analysis of \kpi scattering at the LASS experiment~\cite{Aston:179353}. 

The interference of an S-wave in a predominantly P-wave system has
previously been used to disambiguate otherwise equivalent solutions
for the value of the \CP-violating phase in \Bz~\cite{Aubert:2004cp}
and \Bs~\cite{Aaij:2012eq} oscillations.  In the determination of
$\varphi_s$ in the $\decay{\Bs}{\jpsi\phi}$ decay it was also shown
that it is required to take the S-wave contribution into account
~\cite{Xie:2009fs} and this has subsequently been done for the
experimental measurements~\cite{CDF:2011af,Abazov:2011ry,LHCb:2011aa}.
The interference of a \kpi S-wave in the angular analysis of
\BdToKstmm has previously been considered in
Refs.~\cite{Becirevic:2012dp,Matias:2012qz}.  In both references, the
authors show that the presence of the S-wave can introduce significant
biases to angular observables in the decay.  We extend these studies
to explore the consequences of the S-wave contribution for the present
and future experimental analyses.  Further, we explore the interplay
between statistical and systematical uncertainties for different
analysis approaches.

In this paper, we detail how a generic \kpi S-wave contribution
to \BdToKstll can be included in the angular analysis.  Firstly, we
develop the formalism set out in~\cite{Lu:2011jm} to explicitly
include a spin-0 S-wave and a spin-1 P-wave state in
the \BdToKpill angular distribution. Here \Kstarz is used for any neutral 
kaon state which decays to \kpi. The impact of an S-wave contribution on the 
determination of the theoretical observables is evaluated in two ways:
 in the first we look for the minimum sample
size in which an S-wave contribution (such as measured
in~\cite{Aubert:2004cp}) significantly biases the angular observables; 
secondly we determine, for a given sample size, the minimum S-wave 
contribution
needed to bias the angular observables. We then
demonstrate how the S-wave contribution can be correctly taken into
account and evaluate the effect of this on the statistical precision
that can be obtained on the angular observables with a given number of signal
events.

\section{The \BdToKstll angular distribution}

The differential angular distribution for \BdToKstll is expressed as a
function of the five kinematic variables (\ctl, \ctk, $\phi$, \psq
and \qsq).  The angle
$\theta_K$ is defined as the angle between the \Kp and the \Bd momentum
vector in the rest frame of the \Kstarz.  The angle $\theta_l$ is
similarly defined between the \ellp in the rest frame of the dilepton
pair and the momentum vector of the \Bd.  The angle $\phi$ is defined
as the signed angle between the planes, in the rest frame of the \Bd,
 formed by the dilepton pair and the \kpi pair 
respectively.\footnote{This is the same sign convention for \ctl and \ctk 
as used by the \babar, \belle, \cdf and \lhcb 
experiments~\cite{Aubert:2008ju,PhysRevLett.103.171801,Aaltonen:2011cn,Aaij:2011aa} 
and the same $\phi$ convention as used in 
\lhcb~\cite{LHCb-CONF-2012-008}.}
The mass squared of the \kpi system is denoted \psq and the 
mass squared of the dilepton
pair \qsq.  
The angular distribution is given as a  function
of \ctl, \ctk and $\phi$ as 
\begin{align}
\frac{\text{d}^5\Gamma}{\text{d}\qsq \text{d}p^2 \dctk \dctl \deriv\phi} = & 
\frac{3}{8} \bigg( I_1^c + 2I_1^s + (I_2^c + 2I_2^s) \cos2\theta_l  + 2I_3\stlsq\cos2\phi \nonumber \\
&+ 2\sqrt{2}I_4\sin2\theta_l\cos\phi  + 2\sqrt{2}I_5\stl\cos\phi + 2I_6\ctl  \\ 
&+ 2\sqrt{2}I_7\stl\sin\phi  + 2\sqrt{2}I_8\sin2\theta_l\sin\phi + 2\sqrt{2}I_9\stlsq\sin2\phi \bigg) \nonumber
\end{align}
Ignoring scalar and tensor contributions, the complete set of angular terms are
\begin{align}
I_1^c &= |\mathcal{A}_{0L}|^2 + |\mathcal{A}_{0R}|^2 + 8\frac{m_l^2}{\qsq}\Re\left(\mathcal{A}_{L0}\mathcal{A}_{R0}^{*}\right) + 4\frac{m_l^2}{\qsq}|\mathcal{A}_t|^2   \frac{}{} ,\nonumber   \\
I_1^s &= \frac{3}{4} \left( |\mathcal{A}_{L||}|^2 + |\mathcal{A}_{L\bot}|^2 + (L\to R) \right)  \left( 1 - \frac{4m_l^2}{\qsq} \right) +  \frac{4m_l^2}{\qsq}\Re\left( \mathcal{A}_{L\bot} \mathcal{A}_{R\bot}  + \mathcal{A}_{L||} \mathcal{A}_{R||} \right)  \frac{}{} ,\nonumber \\
I_2^c &= - \beta_l^2 \left( |\mathcal{A}_{L0}|^2 + |\mathcal{A}_{R0}|^2 \right)  \frac{}{} ,\nonumber \\
I_2^s &= \frac{1}{4} \beta_l^2  \left( |\mathcal{A}_{L||}|^2 + |\mathcal{A}_{L\bot}|^2 + |\mathcal{A}_{R||}|^2 + |\mathcal{A}_{R\bot}|^2  \right)  \frac{}{}  ,\nonumber  \\
I_3 &= \frac{1}{2} \beta_l^2  \left( |\mathcal{A}_{L\bot}|^2 - |\mathcal{A}_{L||}|^2 + |\mathcal{A}_{R\bot}|^2 - |\mathcal{A}_{R||}|^2 \right)  \frac{}{} ,\nonumber  \\
I_4 &= \frac{1}{\sqrt{2}}   \beta_l^2  \left(  \Re(\mathcal{A}_{L0}\mathcal{A}_{L||}^{*}) + (L\to R) \right)  \frac{}{} , \\
I_5 &= \sqrt{2}  \beta_l \left(  \Re(\mathcal{A}_{L0}\mathcal{A}_{L\bot}^{*}) - (L\to R) \right)  \frac{}{} ,\nonumber \\
I_6 &= 2   \beta_l \left( \Re(\mathcal{A}_{L||}\mathcal{A}_{L\bot}^{*}) - (L\to R) \right)  \frac{}{} ,\nonumber \\
I_7 &= \sqrt{2}  \beta_l \left(  \Im(\mathcal{A}_{L0}\mathcal{A}_{L||}^{*}) - (L\to R) \right)  \frac{}{} ,\nonumber \\
I_8 &= \frac{1}{\sqrt{2}}  \beta_l^2 \left(  \Im(\mathcal{A}_{L0}\mathcal{A}_{L\bot}^{*}) + (L\to R) \right)  \frac{}{} ,\nonumber \\
I_9 &=  \beta_l^2  \left(  \Im(\mathcal{A}_{L||}\mathcal{A}_{L\bot}^{*}) + (L\to R) \right) \frac{}{} , \nonumber 
\end{align}
where $\mathcal{A}_{H(0,||,\bot,t)}$ are the \Kstarz helicity amplitudes and 
$\beta_l^2 = 1 - 4m_l^2/\qsq$~\cite{PhysRevD.61.114028}. 
 In this paper the lepton mass
is assumed to be insignificant, such that the angular terms with 
$m^2_l/\qsq$ dependence can be neglected and $\beta_l=1$ such that $I_{1}$ and 
$I_2$ can be related by $I_2^c = - I_1^c$ and $I_2^s = \frac{1}{3} I_1^s$.

For a \kpi state which is a combination of different spin states,
the amplitudes for a given handedness ($H=L,R$) can be expressed as a
sum over the resonances ($J$)
\begin{equation}
\label{eq:amp}
\begin{aligned}
\mathcal{A}_{H,0/t}(\psq,\qsq) &= \sum_{J\geq0} \sqrt{N_J} \ M_{J,H,0}(\qsq) \ P_J(p^2) \ Y_J^0 (\theta_K,0) , \\
\mathcal{A}_{H,||/\bot}(\psq,\qsq) &= \sum_{J\geq1} \sqrt{N_J} \ M_{J,H,||/\bot}(\qsq) \ P_J(p^2) \ Y_J^{-1} (\theta_K,0) , 
\end{aligned}
\end{equation}
where $Y_J^m (\theta_K, 0)$ are the spherical harmonics, $M$
is the matrix element and $P_J(\psq)$ is the propagator of
the spin state which encompasses the \psq dependence. 
 A detailed description of the spin-dependent matrix
elements and normalisation factors can be found
in Ref.~\cite{Lu:2011jm}.

\section{Angular distribution of \BdToKpill for a combined S- and P-wave}

For \kpi masses below  
$1200\mev$,\footnote{Natural units are assumed throughout this paper} the
contribution to the amplitudes from the D-wave $\Kstarz(1430)$ is so
small that it can be ignored~\cite{Aston:179353} and only the $J=0,1$ terms in
the sums of Eq.~\ref{eq:amp} will be considered.
The S-wave contribution to these amplitudes only enters in $\mathcal{A}_{0}$ giving 
\begin{align}
\label{eq:amps1}
\mathcal{A}_{H,0} &= \frac{1}{\sqrt{4\pi}} A_{0,H,0} +\sqrt{ \frac{3}{4\pi}} A_{1,H,0} \ctk ,\nonumber\\
\mathcal{A}_{H,||} &= \sqrt{\frac{3}{4\pi}} A_{1,H,||} \ctk  ,  \\
\mathcal{A}_{H,\bot} &= \sqrt{\frac{3}{8\pi}} A_{1,H,\bot} \stk  ,\nonumber 
\end{align}
where the spherical harmonics have been expanded out, leaving the propagator and the matrix element as part of the spin-dependent amplitudes
\begin{equation}
\label{eq:amps2}
\begin{split}
A_{0,H,0} &\propto \  M_{0,H,0} (\qsq) \  P_0(\psq) , \\
A_{1,H,0} &\propto  \  M_{1,H,0} (\qsq)\  P_1(\psq) , \\
A_{1,H,\bot} &\propto  \  M_{1,H,\bot} (\qsq)\  P_1(\psq) , \\
A_{1,H,||} &\propto  \  M_{1,H,||} (\qsq)\ P_1(\psq) , 
\end{split}
\end{equation}
where the first index denotes the spin and the normalisation from the three-body phase space factor is omitted. 
The propagator for the P-wave is described by a relativistic Breit-Wigner distribution with the amplitude given by
\begin{align}
P_1(\psq) = \frac{ m_{\Kstarzo} \Gamma_{\Kstarzo}(\psq)}{ m_{\Kstarzo}^2 - \psq + i \  m_{\Kstarzo} \Gamma_{\Kstarzo}(\psq)} 
\end{align}
where $ m_{\Kstarzo}$ is the resonant mass and 
\begin{align}
\Gamma_{\Kstarzo}(\psq) = \Gamma_{\Kstarzo}^0 \left( \frac{ t }{t_0} \right)^3 \left( \frac{ m_{\Kstarzo} }{ p } \right) \frac{ B\left(tR_P\right) }{ B\left(t_0R_P\right) }
\end{align}
the running width. Here $t$ is the \Kp momentum in the rest frame of the \kpi system and $t_0$ is t evaluated at the \kpi pole mass.
$B$ is the Blatt-Weisskopf damping factor~\cite{Blatt:628052} with a radius $R_P$.
The amplitude can be defined in terms of a phase ($\delta$) through the substitution
\begin{align}
\cot \delta = \frac{   m_{\Kstarzo}^2 - \psq }{  \Gamma_{m_{\Kstarzo}}(\psq) m_{\Kstarzo} }
\end{align}
to give the polar form of the relativistic Breit-Wigner propagator 
\begin{align}
P_1(\psq) =   \frac{1}{\cot\delta - i } 
\end{align}
The LASS parametrisation of the S-wave~\cite{Aston:179353} 
can be used to describe a generic \kpi S-wave.  
In this parametrisation, the S-wave propagator is defined as 
\begin{align}
  \label{eq:LASS1}
P_0(\psq) = \frac{p}{t} \left( \frac{1 }{ \cot\delta_B - i } + e^{2i\delta_B}( \frac{1}{\cot\delta_R - i }) \right)
\end{align}
where the first term is an empirical term from inelastic scattering and the second term is
 the resonant contribution with a phase factor to retain unitarity.
The first phase factor is defined as
\begin{align}
  \label{eq:LASS2}
\cot\delta_B = \frac{1}{ta} + \frac{1}{2}rt ,
\end{align}
where $r$ and $a$ are free parameters and $t$ is defined previously,
while the second phase factor describes the $\Kstarzz(1430)$
 through 
\begin{align}
  \label{eq:LASS3}
\cot\delta_R = \frac{ m_{\mathrm{S}}^2 - \psq }{ \Gamma_{\mathrm{S}}(\psq) m_{\mathrm{S}}  }.
\end{align}
Here, $m_{\mathrm{S}}$ is the S-wave pole mass and $\Gamma_{\mathrm{S}}$ is the 
running width using the pole mass of the $\Kstarzz(1430)$.
The overall strong phase shift between the results from the LASS scattering experiment and 
measured values for \Bd\to\jpsi\kpi has been found to be consistent with $\pi$~\cite{Aubert:2004cp}. 
The parameters for the \psq spectrum used in this paper are given in Table~\ref{tbl:params}.

\begin{table}[tb]
\centering
\caption{Parameters of the \kpi resonances used to generate toy data sets. The \Kstar masses and widths 
are taken from Ref.~\cite{Beringer:1900zz} and the \Kstarzo Blatt-Weisskopf radius and 
the LASS parameters are taken from Ref.~\cite{Aubert:2008aa}~\label{tbl:params}}
\begin{tabular}{|c|c|c|c|c|c|c|}
\hline
State. & mass & $\Gamma$ & R & $r$ & $a$ & $ \delta_{\mathrm{S}} $ \\
  & (\mev) & $(\mev)$ & ($\gev)^{-1}$ & $(\gev)^{-1}$ & $(\gev)^{-1}$ &  \\
\hline
\Kstarzo & $ 894.94  \pm 0.22$ & $ 48.7 \pm 0.8 $ & $3.0$  & \  & \ &\  \\
\Kstarzz & $ 1425  \pm 50 $ & $ 270 \pm 80 $ & $1.0$ & $1.94$ & $1.73$ &  $\pi$ \\
\hline
\end{tabular}
\end{table}

The angular terms modified by the inclusion of the S-wave are $I_{1,2,4,5,7,8}$ and the complete set of angular terms expressed in terms of the spin-dependent amplitudes is
\begin{align}
\label{eq:angularcoeff1}
I_1^c &=  \frac{1}{4\pi} |A_{0L0}|^2 + \frac{3}{4\pi} |A_{1L0}|^2\ctksq + 2 \frac{\sqrt{3}}{4\pi} |A_{0L0}||A_{1L0}|\cos\delta_{0,0}^L \ctk + (L\to R) \frac{}{}  \nonumber \\
I_1^s &= \frac{3}{4} \frac{3}{8\pi} \left( |A_{1L||}|^2 + |A_{1L\bot}|^2 + (L\to R) \right) \frac{}{}\stksq \nonumber   \\
I_2^c &= - I_1^c , \qquad \, I_2^s = \frac{1}{3} I_1^s \nonumber\\ 
I_3 &= \frac{1}{2}  \frac{3}{8\pi}  \left( |A_{1L\bot}|^2 - |A_{1L||}|^2 + (L\to R) \right) \stksq  \frac{}{}\nonumber\\
I_4 &= \frac{1}{\sqrt{2}} \left[  \frac{1}{4\pi}\sqrt{\frac{3}{2}}\Re( A_{0L0}A_{1L||}^{*}) \cos\delta_{0,||}^L \stk  \right. \frac{}{} \nonumber\\
      &+ \left. \frac{3}{4\pi}\sqrt{\frac{1}{2}}\Re( A_{1L0}A_{1L||}^{*})  \stk \ctk  + ( L \to R )  \right] \frac{}{}\nonumber \\
I_5 &= \frac{1}{\sqrt{2}} \left[  \frac{1}{4\pi}\sqrt{\frac{3}{2}}\Re( A_{0L0}A_{1L\bot}^{*}) \cos\delta_{0,\bot}^L \stk  \right. \frac{}{} \nonumber\\
    &+  \left.  \frac{3}{4\pi}\sqrt{\frac{1}{2}}\Re( A_{1L0}A_{1L\bot}^{*})  \stk \ctk  - ( L \to R )  \right] \frac{}{} \\
I_6 &= 2  \frac{3}{8\pi} \left( \Re(A_{1L||}A_{1L\bot}^{*}) - (L\to R) \right) \stksq \frac{}{} \nonumber  \\
I_7 &= \frac{1}{\sqrt{2}} \left[  \frac{1}{4\pi}\sqrt{\frac{3}{2}}\Im( A_{0L0}A_{1L||}^{*}) \cos\delta_{0,||}^L \stk \right. \frac{}{}  \nonumber\\ 
    & + \left.   \frac{3}{4\pi}\sqrt{\frac{1}{2}}\Im( A_{1L0}A_{1L||}^{*})  \stk \ctk   - ( L \to R )  \right] \frac{}{} \nonumber \displaybreak[1]  \\
I_8 &= \frac{1}{\sqrt{2}} \left[  \frac{1}{4\pi}\sqrt{\frac{3}{2}}\Im( A_{0L0}A_{1L\bot}^{*}) \cos\delta_{0,\bot}^L \stk \right. \frac{}{} \nonumber\\
 &  + \left.    \frac{3}{4\pi}\sqrt{\frac{1}{2}}\Im( A_{1L0}A_{1L\bot}^{*})   \stk \ctk  + ( L \to R )  \right]  \frac{}{} \nonumber \\
I_9 &=    \frac{3}{8\pi} \left(  \Im(A_{1L||}A_{1L\bot}^{*}) + (L\to R) \right) \stksq \frac{}{}\nonumber
\end{align}
The interference term of $I_1$ shows how this parametrisation encompasses the strong phase difference between the S and P-wave state. 
The left handed part of the interference term for $I_1$ can be written as 
\begin{align}
2 |A_{0L0}||A_{1L0}|\cos\delta_{0,0}^L \propto 2\, |M_{0,L,0}||P_0(\psq)||M_{1,L,0}||P_1(\psq)| \cos( \delta_{0,0}^L ) 
\end{align}
where 
\begin{align}
 \delta_{0,0}^L =  \delta_{M_{0L0}} + \delta_{P_0} - \delta_{M_{1L0}} - \delta_{P_1}  .
\end{align}
where $\delta_{M_{JL0}}$ is the phase of the  longitudinal matrix element and 
$\delta_{P_J}$ is the phase of the propagator.
The phases in the interference terms for $I_{4,5,7,8}$ can be similarly defined.
For real matrix elements,~i.e.~nearly true in the Standard Model, the phases are 
equal for both handed interference terms $\delta^L = \delta^R$.
The phase difference between the S-wave and the P-wave propagators can be 
expressed as a single strong phase, $\delta_{\mathrm{S}}$.

\begin{figure}[tb]
\centering
\includegraphics[width=0.66\textwidth]{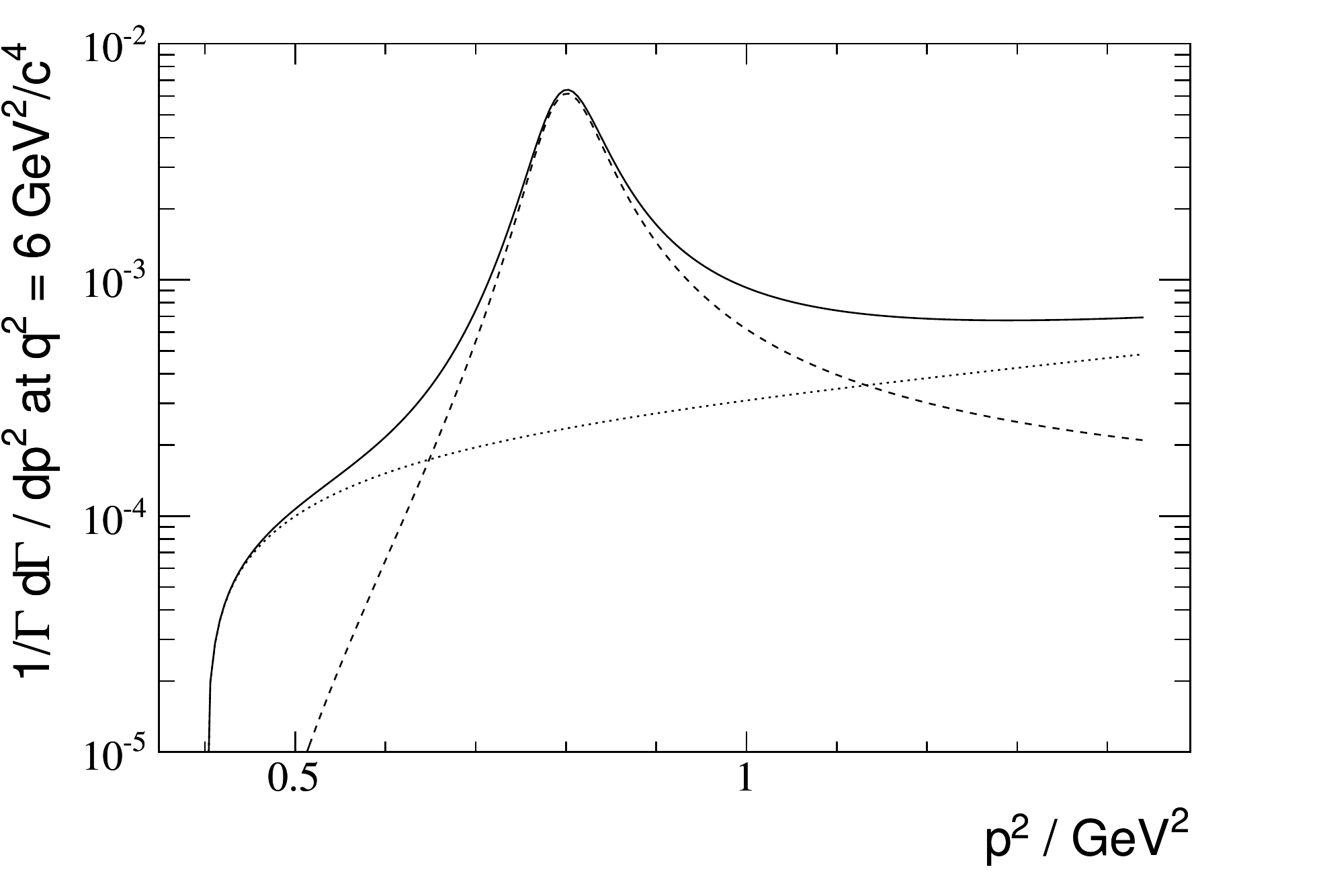}
\caption{ An illustration of the \psq spectrum for the P-wave (dashed) and the S-wave (dash-dotted). 
The total distribution from both states  is the 
solid line. The values were calculated at $q^2 = 6 \gev^2$ by
 integrating out the angular distribution of \BdToKpill using equal 
matrix elements for each state. The S-wave fraction here is 16\% between $800<p<1000 \mev$~\label{fig:psqat6} } 
\end{figure}

The \psq spectrum for the  \BdToKpill angular distribution can
 be calculated by summing over the S- and P-waves and integrating 
out the \ctl, \ctk and $\phi$ dependence. This is illustrated in 
Fig.~\ref{fig:psqat6}
where the matrix elements from 
Refs~\cite{Egede:2008uy,Egede:2010zc} at a \qsq value of $6\gev^2$ are used. 
Here the S-wave amplitude is assumed to be equivalent to the longitudinal P-wave amplitude.
The S-wave fraction in the $800 < p < 1000\mev$ window around around the P-wave 
is calculated to be 16\% when using this approximation.
As will be seen later there are no interference terms left in the angular distribution after the integral
 over \ctk.

\section{The effect on \BdToKpill observables}

So far the forward-backward asymmetry (\AFB), the fraction of the \Kstarz 
longitudinal polarisation (\FL ) and two combinations of the transverse 
amplitudes (\AT2 and \AIm ) have been measured.
As such, these are the observables that will be concentrated on here.

\AFB  is defined in terms of the amplitudes as
\begin{align}
\label{eq:theoafb}
\AFB(\qsq) &= \frac{3}{2} \frac{  \Re(A_{1L||}A_{1L\bot}^{*}) - 
\Re(A_{1R||}A_{1R\bot}^{*})}{|A_{10}|^2 + |A_{1||}|^2 + |A_{1\bot}|^2 }
\end{align}
for a pure P-wave state where the generic combination of amplitudes 
    $A_{Ji}A_{Ji}^{*}$ is defined as
\begin{align}
\label{eq:ampsq}
A_{Ji} A_{Ji}^{*} &= A_{JiL} A_{JiL}^{*} + A_{JiR} A_{JiR}^{*}.
\end{align}
where $ i \in \{0,||,\bot,t\}$ and $J=0,1$.
The factorisation of the amplitudes into matrix elements and the 
propagators removes the \psq dependence from the theoretical observables. 
In a similar way, \FL, \AT2 and \AIm are defined as
\begin{equation}
\begin{split}
\FL(\qsq) &= \frac{  |A_{10}|^2 } {|A_{10}|^2 + |A_{1||}|^2 + |A_{1\bot}|^2} ,\\
\AT2 (\qsq) &=  \frac{ |A_{1\bot}|^2 - |A_{1||}|^2  }{ |A_{1\bot}|^2 + |A_{1||}|^2 } = ( 1 - \FL ) \frac{ |A_{1\bot}|^2 - |A_{1||}|^2 }{ |A_{10}|^2 + |A_{1||}|^2 + |A_{1\bot}|^2} ,\\
\AIm(\qsq) &= \frac{  \Im(A_{1L||}A_{1L\bot}^{*})+\Im(A_{1R||}A_{1R\bot}^{*})}{|A_{10}|^2 + |A_{1||}|^2 + |A_{1\bot}|^2} .
\end{split}
\end{equation}
These theoretical observables are normalised to the sum of the spin-1 amplitudes.
In terms of the angular distribution, \AFB can also be expressed 
as the difference 
between the number of `forward-going' \mup and the number of 
`backward-going' \mup in the rest frame of the \Bd,
\begin{align}
\label{eq:expafb}
\left[ \int_0^1 - \int_{-1}^0 \right]  \text{d}\ctl
\frac{\text{d}\Gamma}{\text{d}\qsq \text{d}\ctl} / 
\frac{\text{d}\Gamma}{\text{d}\qsq}
\end{align}
which explains the name of the observable. In Ref~\cite{LHCb-CONF-2012-008}, 
this expression was used  to determine the zero-crossing point of \AFB.
The inclusion of the S-wave in the complete angular distribution means 
that \AFB can no longer be determined by 
experimentally counting the number of events with forward-going and  
backward-going leptons, as Eqs.~\ref{eq:theoafb} and~\ref{eq:expafb} 
are no longer equivalent. However, as the S-wave has no forward-backward 
asymmetry, no bias occurs in the determination of the zero-crossing point by  
ignoring the S-wave. 
The total normalisation for the angular distribution changes to the 
sum of S- and P-wave amplitudes,
\begin{align}
\Gamma^{''}  &\equiv \frac{\deriv^2\Gamma}{\deriv\psq\deriv\qsq} =   |A_{10}|^2 + |A_{1||}|^2 + |A_{1\bot}|^2 + |A_{00}|^2 .
\end{align}
such that there is a factor of 
\begin{align} 
\FPi(\psq,\qsq) &= \left( \frac{|A_{10}|^2 + |A_{1||}|^2 + |A_{1\bot}|^2 }
{|A_{10}|^2 + |A_{1||}|^2 + |A_{1\bot}|^2 + |A_{00}|^2} \right)
\end{align}
between the pure P-wave and the admixture of the S and the P-wave. This is the fraction of the yield coming 
from the P-wave at a given value of \psq and \qsq. 
Similarly, the S-wave fraction is defined as 
\begin{align}
\FSi(p^2,q^2) &= \left( \frac{|A_{00}|^2}{|A_{10}|^2 + |A_{1||}|^2 + |A_{1\bot}|^2 + |A_{00}|^2} \right)
\end{align}
and the interference between the S-wave and the P-wave as 
\begin{align}
\ASi(\psq,\qsq)  &= \frac{\sqrt{3}}{2} \left( \frac{ |A_{0L0}||A_{1L0}|\cos\delta_L
+ (L\to R)}{|A_{10}|^2 + |A_{1||}|^2 + |A_{1\bot}|^2 + |A_{00}|^2}\right)
\end{align}
Substituting the above observables into the angular terms gives 
\begin{align}
\label{eq:angularcoeffwithobs}
\frac{I_1^c}{\Gamma^{''}} &=  \frac{1}{4\pi} \FSi + \frac{3}{4\pi} \FPi \FL \ctksq +  \frac{3}{4\pi} \ASi \ctk  ,\nonumber \\
\frac{I_1^s}{\Gamma^{''}} &= \frac{3}{4} \frac{3}{8\pi} \FPi \left( 1 - \FL \right)  \left(1-\ctksq\right)   ,\nonumber \\
\frac{I_2^c}{\Gamma^{''}} &= -  \left( \frac{1}{4\pi} \right. \FSi + \frac{3}{4\pi} \FPi\left( 1 - \FL \right) \ctksq +\left.   \frac{3}{4\pi} \ASi \ctk \ctk \right)  ,\nonumber \\
\frac{I_2^s}{\Gamma^{''}} &= \frac{1}{4}   \frac{3}{8\pi}  \FPi\left( 1 - \FL \right)  \left(1-\ctksq\right)   , \\
\frac{I_3}{\Gamma^{''}} &= \frac{1}{2}  \frac{3}{8\pi}  \FPi  \AT2   \left(1-\ctksq\right)   ,\nonumber \\
\frac{I_6}{\Gamma^{''}} &= 2  \frac{3}{8\pi} \frac{4}{3} \FPi  \AFB   \left(1-\ctksq\right)   ,\nonumber \\
\frac{I_9}{\Gamma^{''}} &=    \frac{3}{8\pi}  \FPi \AIm \left(1-\ctksq\right)    .\nonumber 
\end{align}
 For the purpose of this paper, a
simplification of the angular distribution can be achieved by folding
 the distribution in $\phi$ such that $\phiprime = \phi - \pi $ 
for $\phi < 0 $~\cite{Ksteepubnote}.
The $I_{4,5,7,8}$ angular terms which are dependent on $\cos\phi$ or
$\sin\phi$ are cancelled leaving $I_{1,2,3,6,9}$ in the angular distribution:
\begin{align}
\label{eq:folded}
\frac{\text{d}^5\Gamma}{\text{d}q^2 \text{d}p^2 \dctk \dctl
\text{d}\phiprime} = & \frac{3}{8} \left( I_1^c + 2I_1^s + (I_2^c + 2I_2^s)
\cos2\theta_l  + 2I_3\stlsq\cos2\phiprime \right. \nonumber \\
& \left. + 2I_6\ctl + 2\sqrt{2}I_9\stlsq\sin2\phiprime \frac{}{} \right) .
\end{align}

Combining Equation~\ref{eq:folded} with~\ref{eq:angularcoeffwithobs} gives the differential decay distribution,
\begin{equation}
\label{eq:theo5d}
\begin{split}
\frac{1}{\Gamma^{''}} \frac{\text{d}^5\Gamma}{\text{d}\qsq\text{d}\psq \dctk
\dctl \text{d}\phiprime} =   \frac{9}{16\pi}   \Bigg( & \left(  \frac{2}{3}\FSi  +  \frac{4}{3} \ASi \ctk \right) ( 1 - \ctlsq )    \\
& \ \ \ \ + \FPi  \bigg[\xspace 2 \FL \ctksq ( 1 - \ctlsq )   \\ 
& \ \ \ \ \ \ \ \ + \frac{1}{2} ( 1 - \FL ) ( 1 - \ctksq ) ( 1 + \ctlsq )  \\
& \ \ \ \ \ \ \ \ + \frac{1}{2} ( 1 - \FL ) \AT2 ( 1 - \ctksq ) (  1 - \ctlsq )
\cos2\phiprime  \\
& \ \ \ \ \ \ \ \ +  \frac{4}{3} \AFB ( 1 - \ctksq ) \ctl  \\ 
& \ \ \ \ \ \ \ \ + \AIm ( 1 - \ctksq ) ( 1 - \ctlsq) \sin2\phiprime  \bigg]  \Bigg) . 
\end{split}
\end{equation}
The angular distribution as a function of \ctl  and \ctk is given by integrating over $\phi$ in Eq.~\ref{eq:theo5d}
\begin{equation}
\label{eq:theo4d}
\begin{split}
\frac{1}{\Gamma^{''}} \frac{\text{d}^4\Gamma}{\text{d}\qsq\text{d}\psq\text{d}
\ctk \text{d}\ctl} =   \frac{9}{16} & \Bigg( \left( \frac{2}{3} \FSi + \frac{4}{3} \ASi \ctk \right) ( 1 - \ctlsq )  \\
&+ \FPi  \bigg[  2 \FL \ctksq ( 1 - \ctlsq )  \\ 
& \ \ \ \ \ +\frac{1}{2}  ( 1 - \FL ) ( 1 - \ctksq ) ( 1 + \ctlsq )  \\
& \ \ \ \ \ + \frac{4}{3} \AFB ( 1 - \ctksq ) \ctl   \bigg]  \Bigg) 
\end{split}
\end{equation}
and further integration from Equation~\ref{eq:theo5d} yields the angular distribution for each of the angles,
\begin{equation}
\label{eq:theo3d}
\begin{split}
\frac{1}{\Gamma^{''}} \frac{\text{d}^3\Gamma}{\text{d}\qsq\text{d}\psq\dctl}
&=  \frac{3}{4} \FSi ( 1 - \ctlsq ) + \FPi  \bigg[  \frac{3}{4} \FL ( 1 - \ctlsq ) \\
&  + \frac{3}{8} ( 1 - \FL ) ( 1 + \ctlsq ) + \AFB \ctl  \bigg]  , \\
\frac{1}{\Gamma^{''}} \frac{\text{d}^3\Gamma}{\text{d}\qsq\text{d}\psq\dctk 
} &=   \frac{1}{2} \FSi + \ASi \ctk \\&
+ \FPi \bigg[ \frac{3}{2}  \FL \ctksq +  \frac{3}{4} ( 1 - \FL ) ( 1 - \ctksq ) \bigg] 	, \\
\frac{1}{\Gamma^{''}} \frac{\text{d}^3\Gamma}{\text{d}\qsq\text{d}\psq\text{d}\phiprime
} &= \frac{1}{\pi} \Bigg(  1 + \frac{3}{4} \FSi    + \FPi \bigg[\xspace   \FL  
+ \frac{1}{2} ( 1 - \FL ) \AT2  \cos2\phiprime +  \AIm  \sin2\phiprime  \bigg]  \Bigg) .
\end{split}
\end{equation}
The angular distribution can be integrated over \psq using the weighted integral 
\begin{align}
\label{eq:obsint}
O(\qsq) &=  \frac{ \int \mathcal{O}(\psq,\qsq) \frac{\text{d}^2\Gamma}
{\text{d}\psq\text{d}\qsq} \text{d}\psq }{ \int  \frac{\text{d}^2\Gamma}
{\text{d}\psq\text{d}\qsq} \text{d}\psq }
\end{align}
for the value of the observables integrated over a given region in \psq. 
This leads to the integrated observables \FP, \FS and \AS which are solely 
dependant on \qsq. By definition, the fraction of the S-wave and the P-wave
 sum to one, $\FS + \FP = 1$.
The complete angular distribution without any \psq dependence is given by
\begin{align}
\label{eq:theo4dint}
\frac{1}{\Gamma^{'}} \frac{\text{d}^5\Gamma}{\text{d}\qsq\dctk
\dctl \text{d}\phiprime} =   \frac{9}{16\pi}  & \Bigg( \left( \frac{2}{3} \FS +  \frac{4}{3} \AS \ctk \right) ( 1 - \ctlsq )   \nonumber \\
&+ (1-\FS)  \bigg[\xspace 2 \FL \ctksq ( 1 - \ctlsq )  \nonumber \\ 
&  \ \ \ \ \ \ \ \ \ \ \ \  + \frac{1}{2} ( 1 - \FL ) ( 1 - \ctksq ) ( 1 + \ctlsq )  \nonumber \\
&  \ \ \ \ \ \ \ \ \ \ \ \  + \frac{1}{2} ( 1 - \FL ) \AT2 ( 1 - \ctksq ) (  1 - \ctlsq ) \cos2\phiprime  \nonumber \\
&  \ \ \ \ \ \ \ \ \ \ \ \  +  \frac{4}{3} \AFB ( 1 - \ctksq ) \ctl  \nonumber \\ 
&  \ \ \ \ \ \ \ \ \ \ \ \  + \AIm ( 1 - \ctksq ) ( 1 - \ctlsq) \sin2\phiprime  \bigg] \Bigg) . 
\end{align}
 where the normalisation of the angular distribution is given by 
\begin{align}
\Gamma^{'} = \frac{\deriv\Gamma}{\deriv\qsq}
\end{align}
The `dilution' effect of the S-wave can clearly be seen from the factor of (1-\FS) that appears in front of the  observables in Eq.~\ref{eq:theo4dint}.

\begin{figure}[tp]
\centering
\hspace{-0.2in}
\subfigure[]{\includegraphics[width=0.48\textwidth]{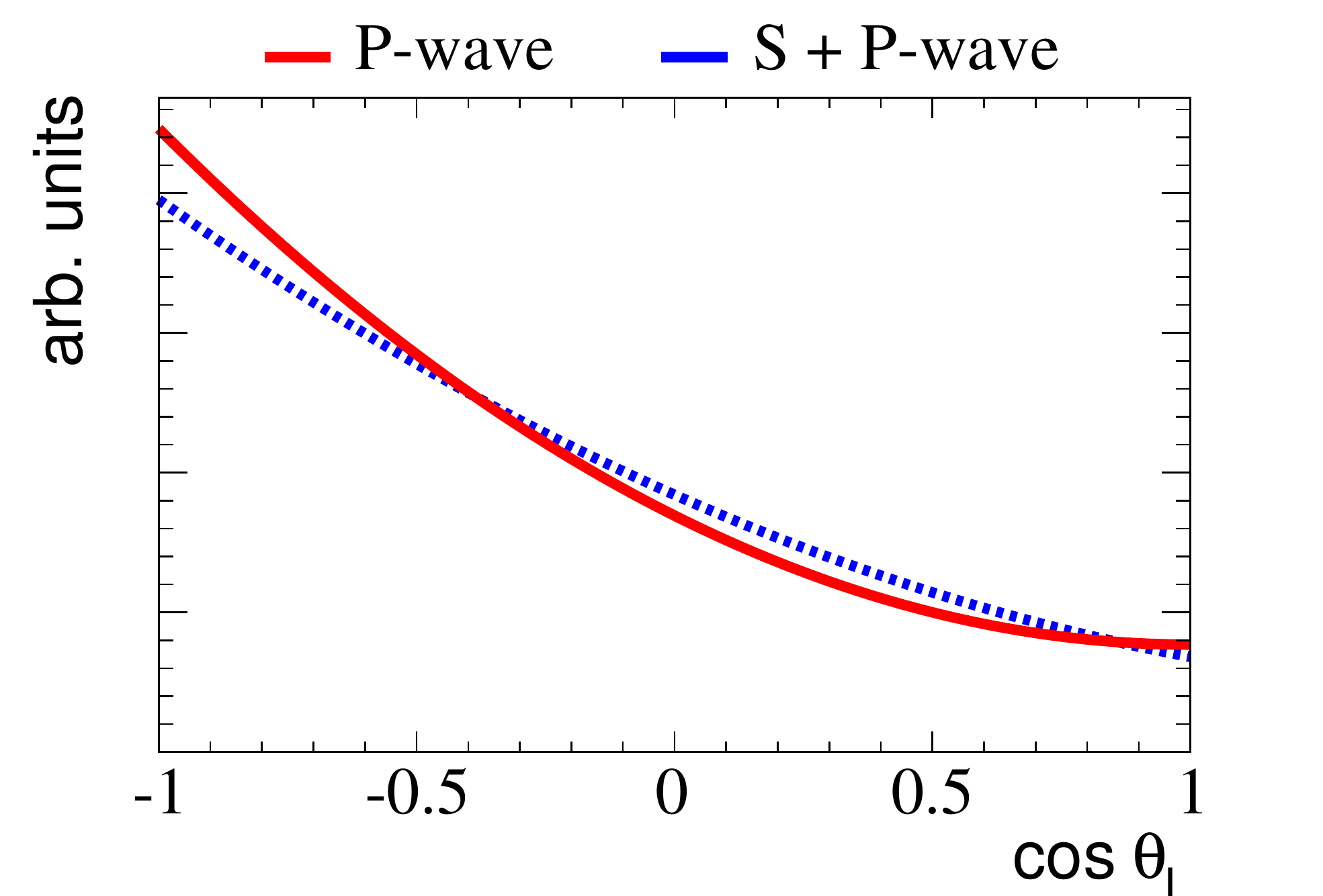}}
\subfigure[]{\includegraphics[width=0.48\textwidth]{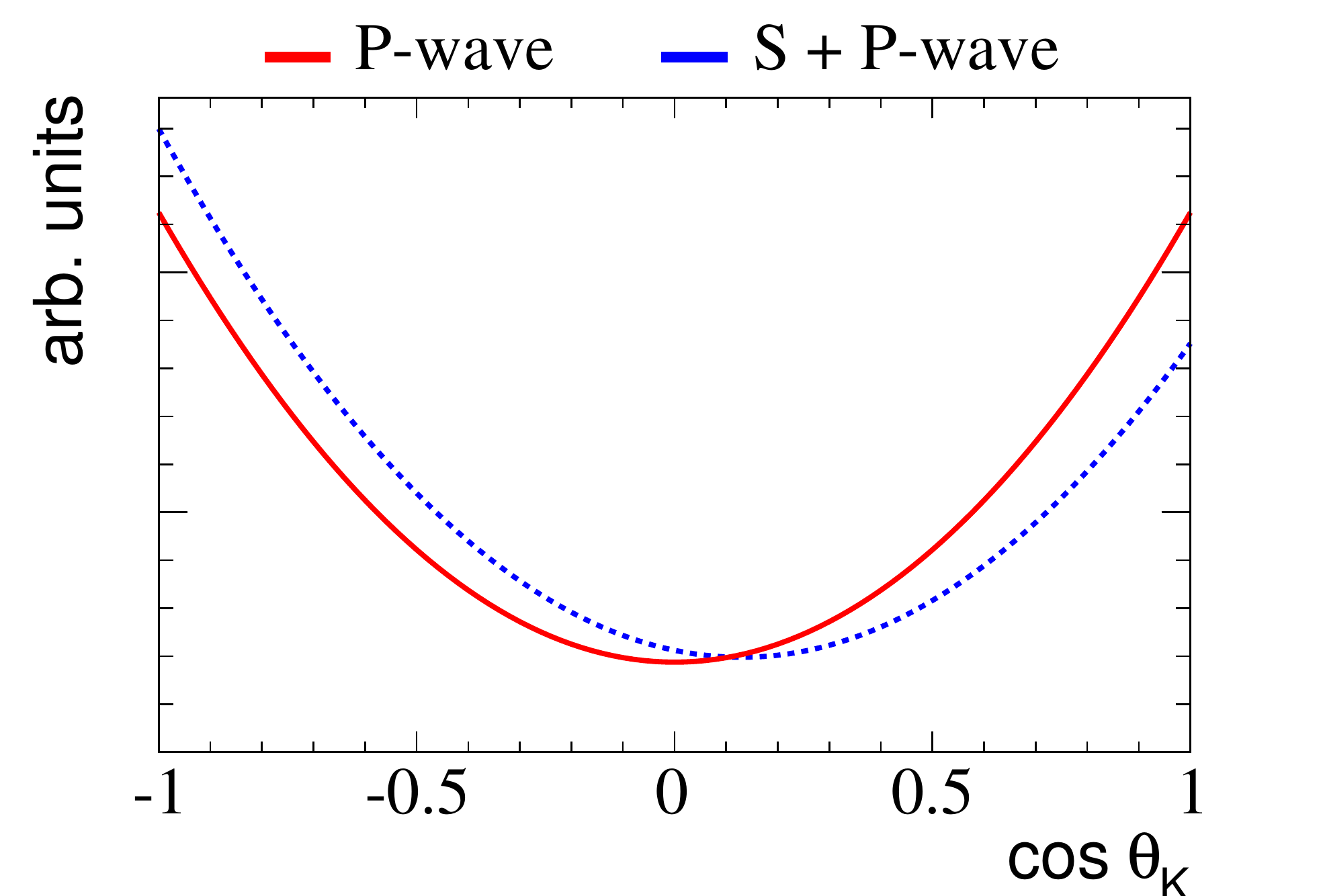}}
\subfigure[]{\includegraphics[width=0.48\textwidth]{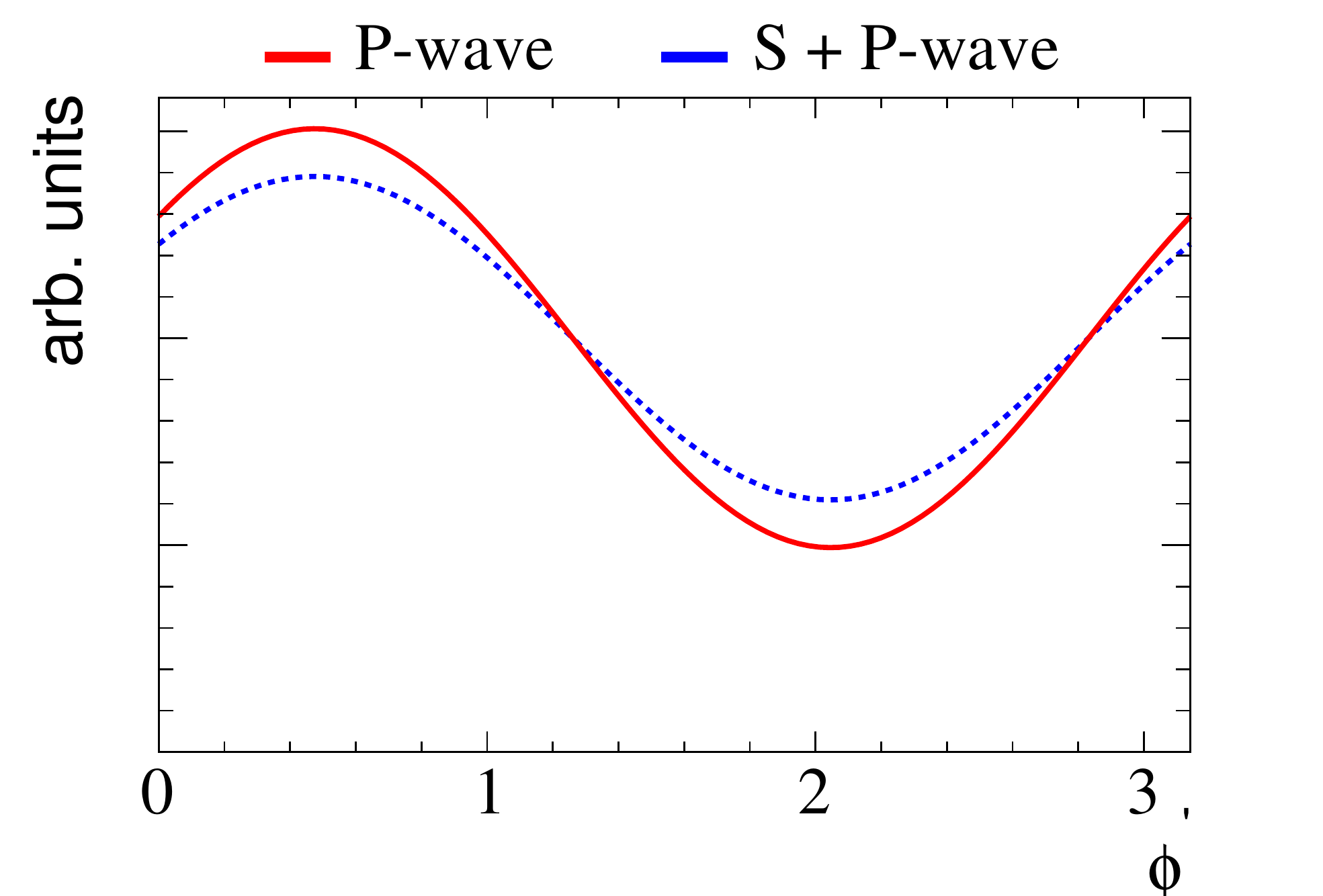}}
\caption{One-dimensional projections of (a) \ctl, (b) \ctk, (c) \phiprime 
for the angular distribution of \BdToKstll with (blue-dashed) and without (red-solid) an 
S-wave component of 7\%. The dilution effect of the S-wave on the asymmetry 
in \ctl and the asymmetric effect in \ctk can be clearly seen.~\label{fig:1dcalc}}
\end{figure}

The effect of an S-wave on the angular distribution as a function of \ctk, \ctl and \phiprime as illustrated in Figure~\ref{fig:1dcalc}. 
Here it is possible to see that the asymmetry in \ctl, given by \AFB, 
has decreased and that there is an asymmetry in \ctk introduced by the interference term.

\section{Effect of an S-wave on the angular analysis}

In an angular analysis of \BdToKpill, the S-wave can be considered to be 
a systematic effect that could bias the results of the angular observables.
The implications of this systematic effect are tested by generating toy 
Monte Carlo experiments and fitting the angular distribution to them.
The results of the fit to the observables are evaluated for multiple toy 
datasets.

The effect of the S-wave is evaluated for two different cases.
Firstly, the effect of S-wave interference  is examined as a function 
of the size of the dataset used.
The aim of this is to give an idea of the current situation and the
 possible implications on future measurements of \BdToKpill.
Datasets of sizes between 50 and 1000 events are tested.
For comparison, the latest results from LHCb~\cite{LHCb-CONF-2012-008} 
have between 20 and 200 signal events in the 6 different \qsq bins considered.
Secondly, the effect of different levels of S-wave contribution is examined. 
At present, the only information about the S-wave fraction is 
the measurement of \FS of approximately 7\% in the decay $\Bd\to\jpsi\kpi$
from~\cite{Aubert:2004cp} for the range $800 < p < 1000 \mev$. 
As the value may be different in \BdToKpill, we consider 
values of \FS in this region ranging from 1\% to 40\%. 
The fraction of the S-wave, \FS, is expected to have some \qsq dependence
 because of the \qsq dependence of the transverse P-wave amplitudes.

The parameters used to generate the toy datasets are summarised in 
Tables~\ref{tbl:params} and~\ref{tbl:obs}.
The values of the angular observables used  to generate toy Monte Carlo
 simulations are taken from the \lhcb angular analysis of 
\BdToKstmm in the $1<\qsq<6 \gev^2$ bin~\cite{LHCb-CONF-2012-008}.
Within errors, these measurements are compatible with the Standard Model prediction for \BdToKstll 
and the central value of the measurement is used.
The nominal magnitude  and phase difference of the S-wave contribution 
are taken from the angular analysis of \Bd\to\jpsi\kpi~\cite{Aubert:2004cp}.

\begin{table}[tb]
\caption{Parameters used to generate toy datasets. \AFB, \FL, \AT2 and \AIm
 are taken from Ref.~\cite{LHCb-CONF-2012-008}  in the 
 $1 < \qsq < 6~(\gev^2)$ bin. 
The \FS value is taken from Ref.~\cite{Aubert:2004cp}~\label{tbl:obs}}.
\centering
\begin{tabular}{|c|c|c|c|c|c|}
\hline
Obs.  & \AFB & \FL & \AT2 & \AIm  & \FS  \\
\hline
Value &$ -0.18 $ &$ 0.66$  & $ 0.294$ & $0.07$  & $0.07$  \\
\hline
\end{tabular}
\end{table}

\begin{figure}[tb]
\centering
\subfigure{\includegraphics[width=0.48\textwidth]{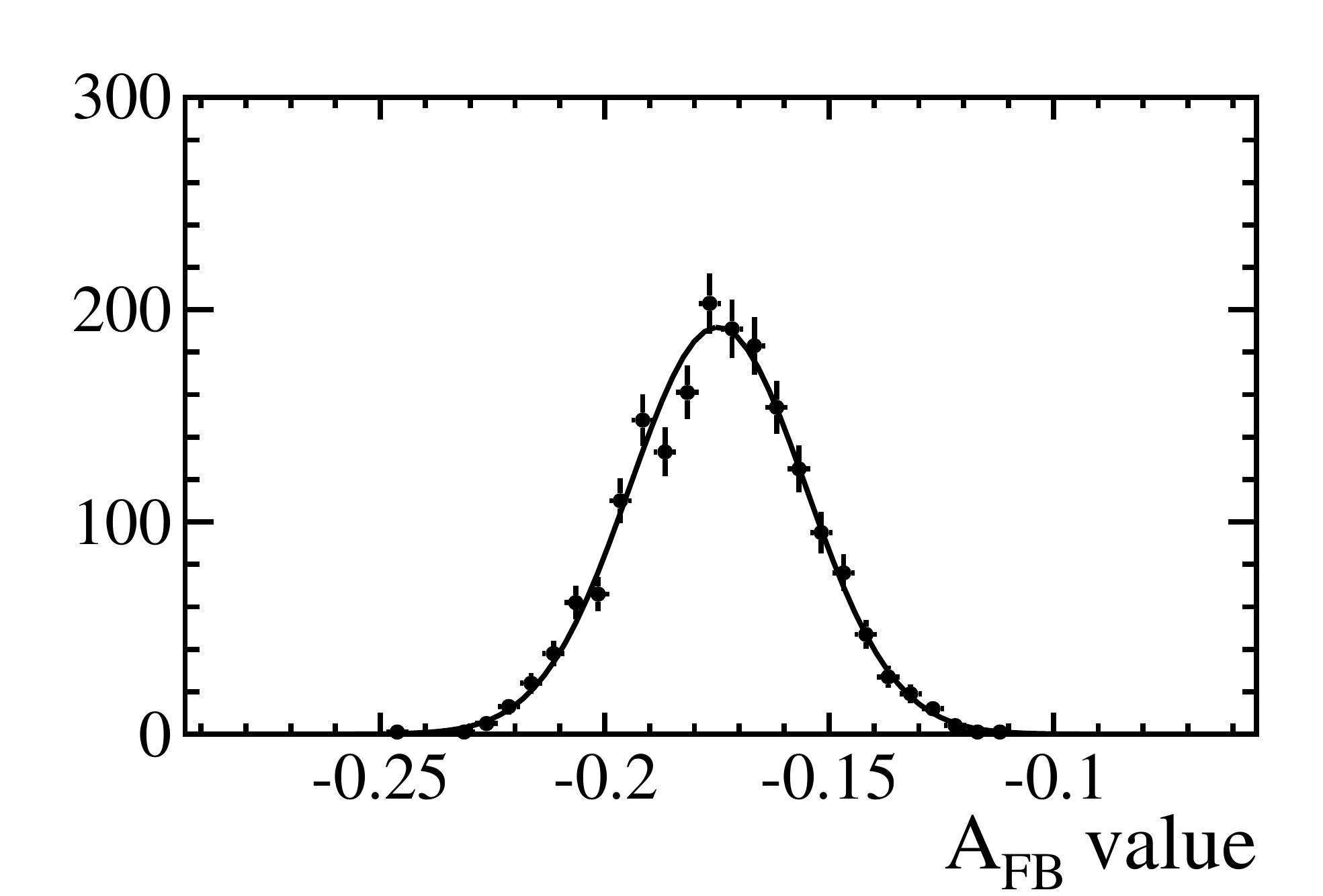}}
\subfigure{\includegraphics[width=0.48\textwidth]{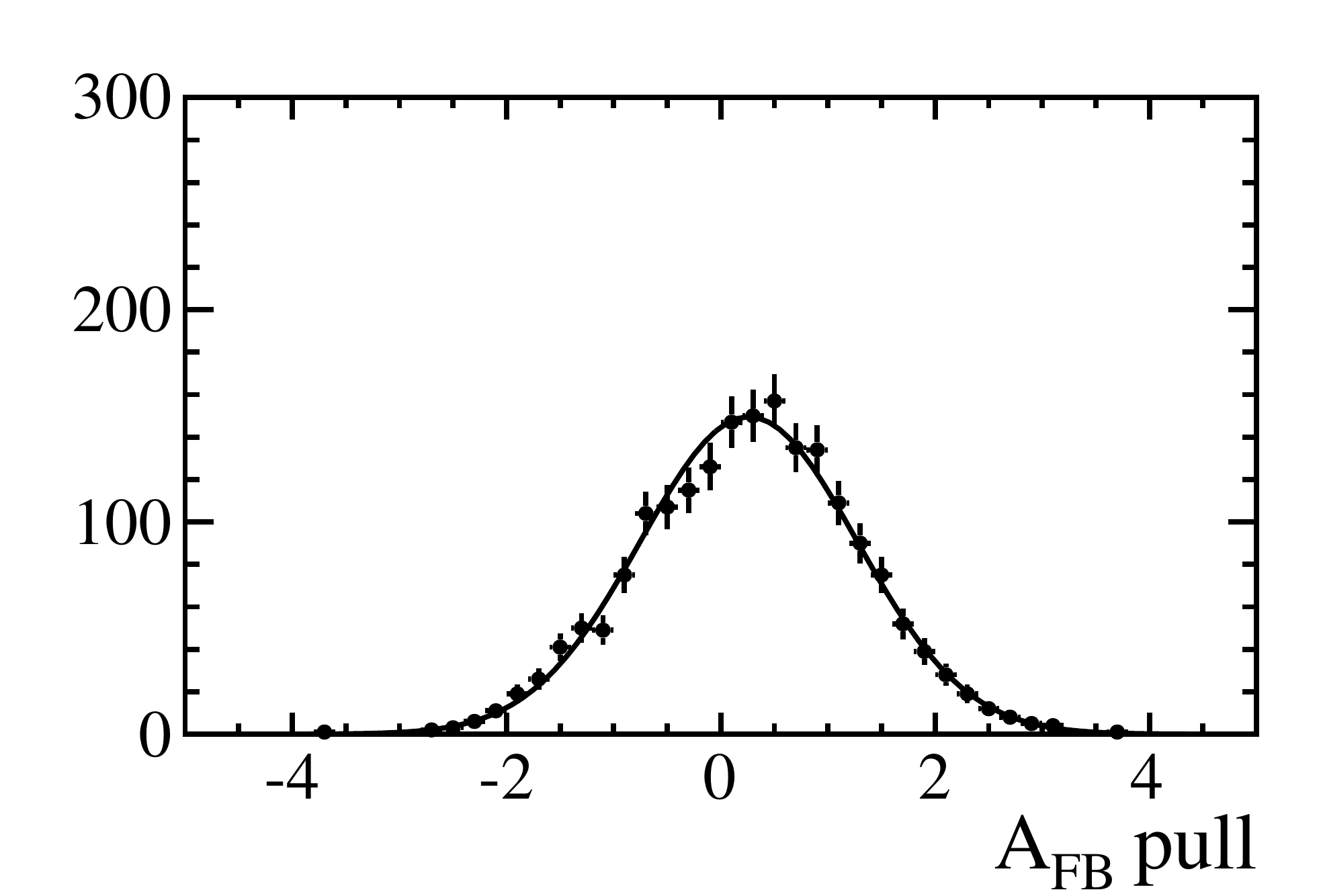}}
\caption{Distribution of (left) the \AFB results and (right) pull 
values for fits to 1000 toy simulations each containing 1000 events.
The S-wave is ignored in these fits. The resolution obtained is 
$(0.026\pm0.001)$. Since the S-wave is ignored there is a non-zero 
pull mean at $(0.26\pm0.02)\sigma$ . The width of the pull 
distribution is consistent with unity at 
 $(1.01\pm0.01)\sigma$.~\label{fig:toyexample}}
\end{figure}

The toy datasets are generated as a function of the
 \ctl, \ctk, $\phi$ and \psq using the angular distribution given in 
 Eq.~\ref{eq:theo5d}.
For each set of input parameters 1000 toy datasets were generated.
For each of these toy datasets, an unbinned log likelihood fit is 
performed that returns the best fit value of the observables 
and an estimate of their error.
The expected experimental resolution is obtained by plotting 
the best fit values of an observable for the ensemble of toy 
simulations as illustrated for \AFB in Fig.~\ref{fig:toyexample} (left)
 The pull value for an observable ($O$) is defined as 
\begin{align}
p_{O}^i= \frac{ O_{\text{fit}}^i - O_{\text{gen}}^i }{ \sigma_{O}^i }
\end{align}
where $\sigma_O^i$ is the estimated error on the fit to the observable $O^i$.
This distribution is seen in Fig.~\ref{fig:toyexample} (right). 
The mean and the width are extracted from a Gaussian fit.
For a well performing fit without bias, the pull distribution should 
have zero mean and unit width. 
A negative pull value implies that the result is underestimated 
and a positive pull value implies overestimation of the true observable.

\subsection{The impact of ignoring the S-wave in an angular analysis of \BdToKstll}

\begin{figure}[tb]
\centering
\subfigure{\includegraphics[width=0.48\textwidth]{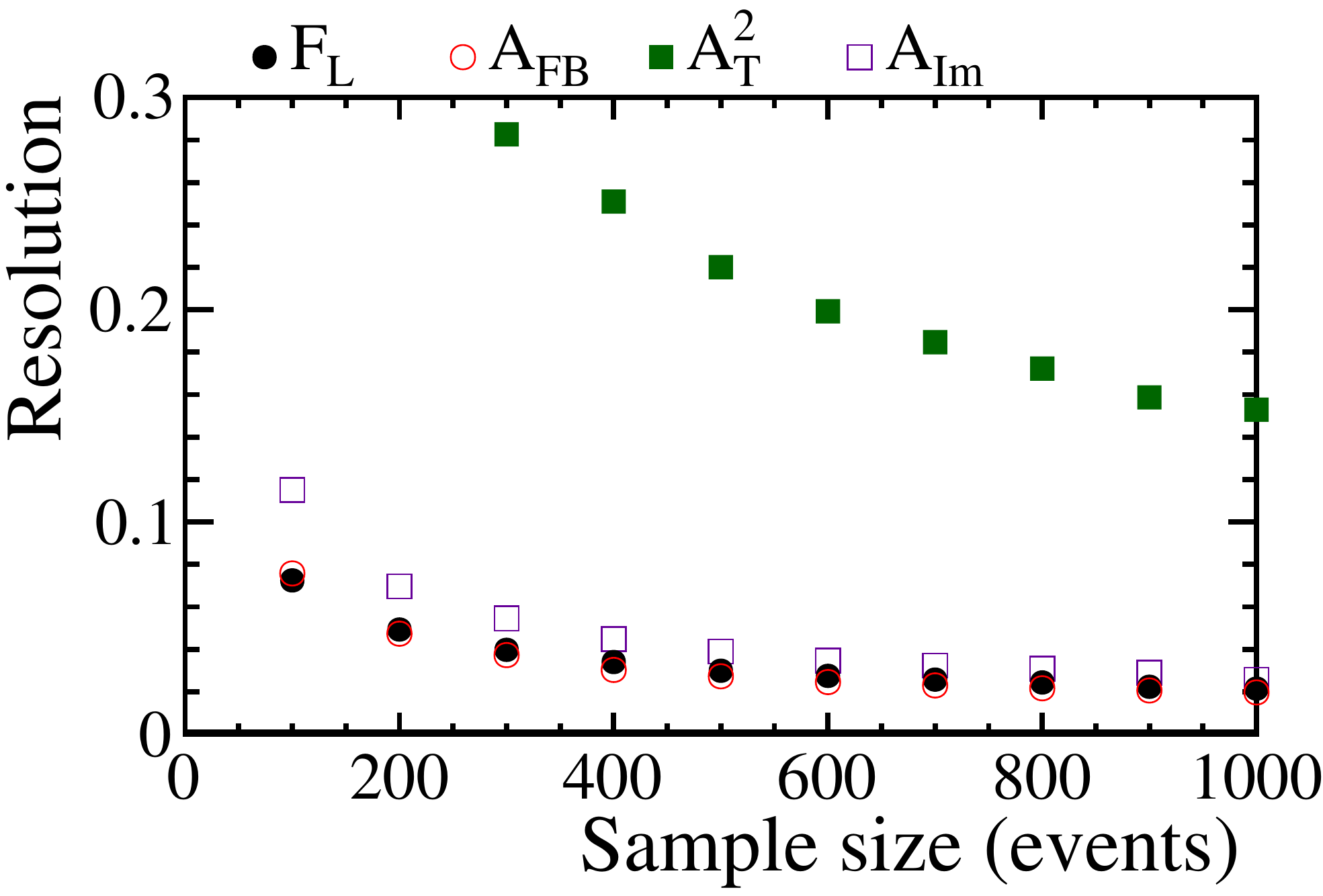}}
\subfigure{\includegraphics[width=0.48\textwidth]{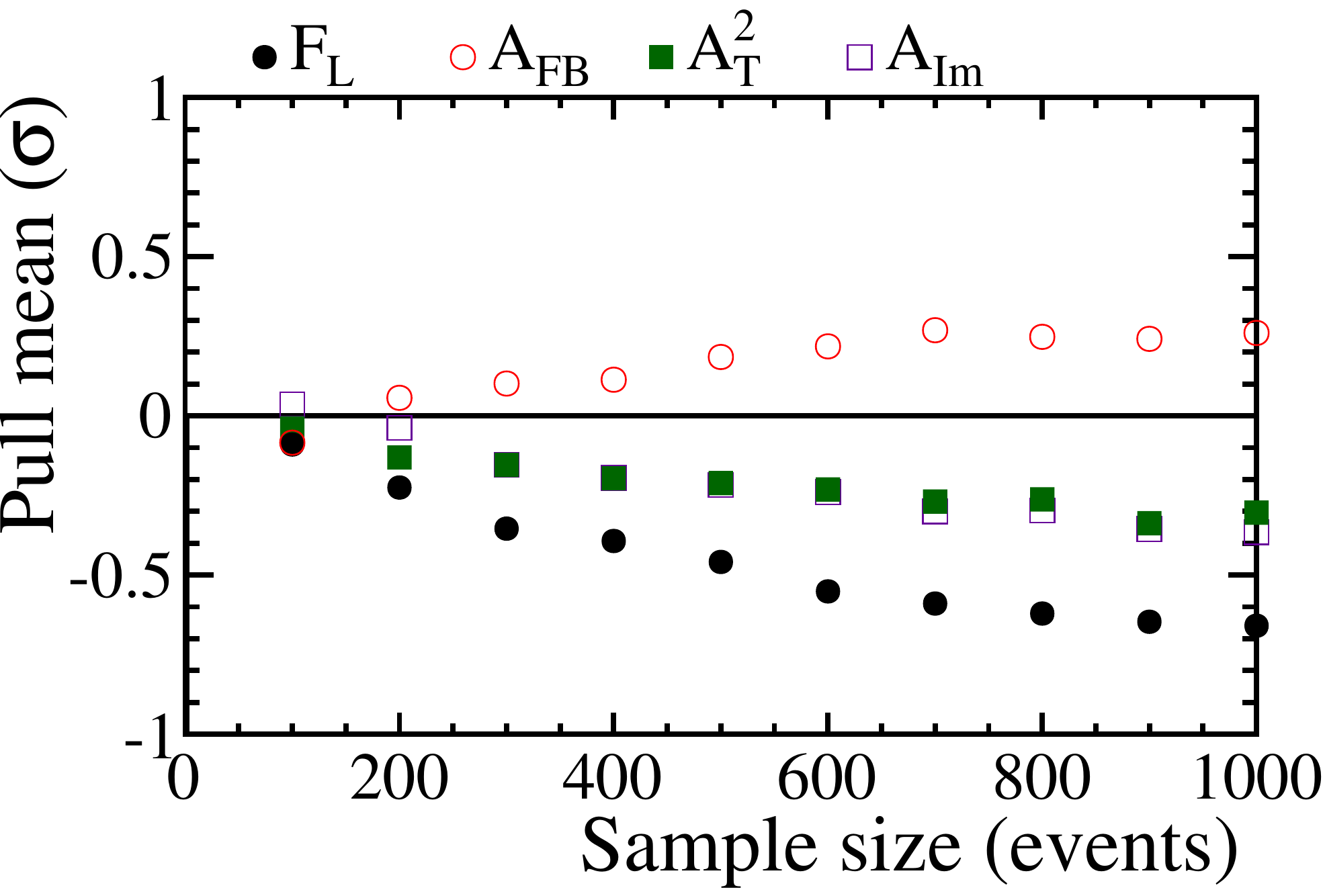}}
\caption{Resolution (left) and pull mean (right)  of 
1000 toy datasets analysed as a pure P-wave state as a 
function of dataset size. It can be seen that the bias 
on the observable increases dramatically as the sample 
size increases. This is because the statistical error 
decreases increasing the sensitivity to the S-wave 
contribution. The bias of \AFB is positive  
because \AFB in negative in the \qsq bin chosen. 
~\label{fig:bias}}
\end{figure}

Firstly, the effect of an S-wave was tested as a function of dataset 
 size in order to find a minimum dataset at which the bias 
 from the S-wave in the angular observables becomes significant.
Datasets were generated for sample sizes ranging from 50 and 
1000 events and analysed assuming a pure P-wave state. 
The results are shown in Fig.~\ref{fig:bias}. 

From Eq.~\ref{eq:theo4d}, it can be seen that \AT2 
has a factor of (1-\FL) in front of it. The large 
value of \FL used in generated the datasets is 
in turn causing \AT2 to have a much worse 
resolution than \AFB, \FL and \AIm. There is 
significant bias (non-zero mean) of the pull distribution for all
 observables when the S-wave is ignored for datasets 
 of more than 200 events. This corresponds to a 
 change of 0.2$\sigma$ in \FL for a dataset of 200 events.
The behaviour can be understood in terms of the $(1-\FS)$ 
factor in Eq~\ref{eq:theo4d}. It gives an offset to the 
fitted value of the observables which are proportional 
to the value of \FS.

\begin{figure}[tb]
\centering
\subfigure{\includegraphics[width=0.48\textwidth]{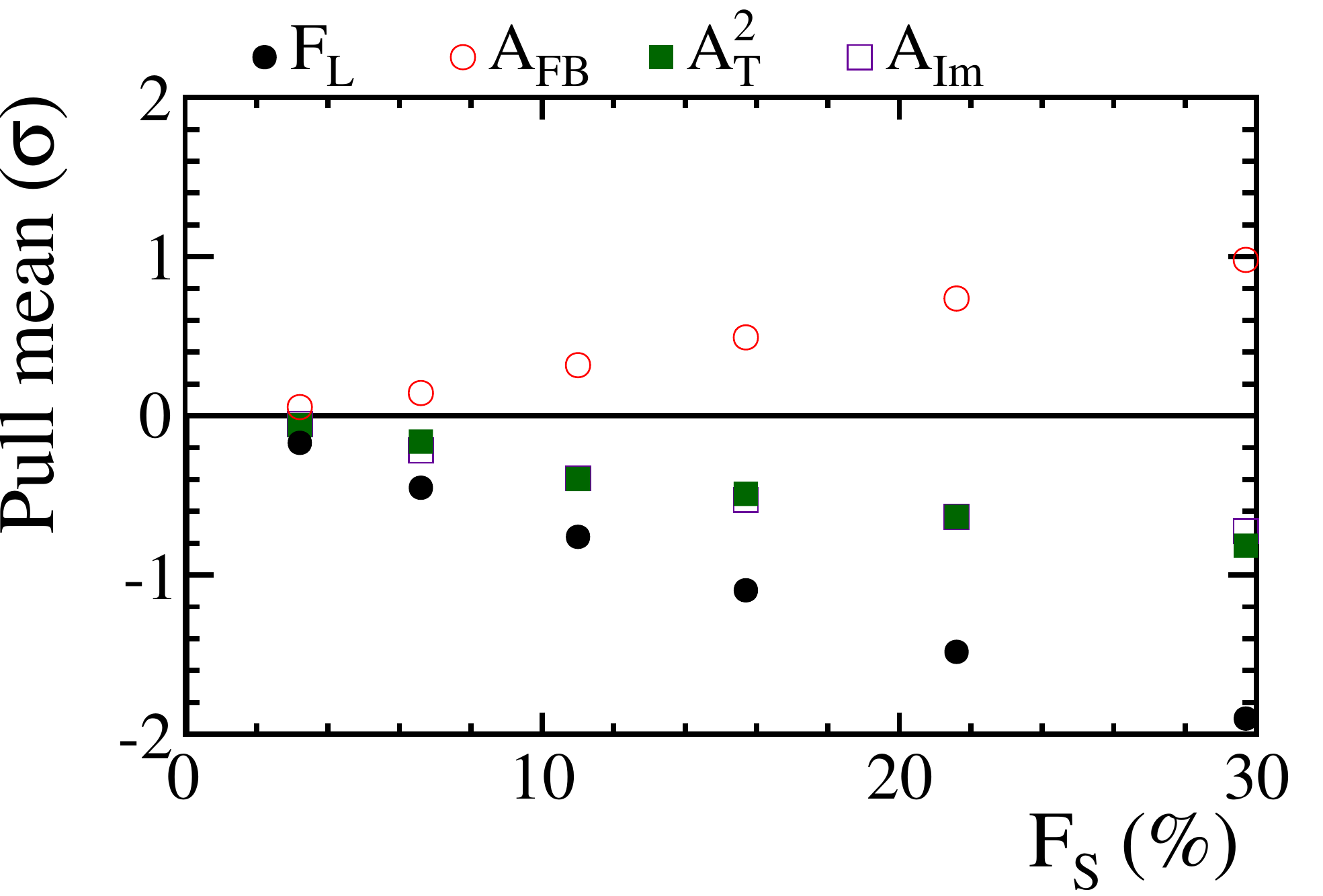}}
\caption{Mean of the pull distribution 
of 1000 toy datasets analysed as a pure 
P-wave state as a function of S-wave contribution. 
The bias can be seen to increase with the size of the S-wave 
contribution in a linear fashion. ~\label{fig:fsval}}
\end{figure}

Secondly, the angular fit was performed on toy 
datasets with an increasing S-wave contribution.
Datasets of 500 events were generated with a 
varying S-wave contribution in the narrow \psq
 mass window of ($800 < p < 1000 \mev$) 
 from no S-wave up to a \FS value of $0.6$. 
The resolution, the mean and width of the pull 
distribution for each of the four observables
 (\AFB, \FL, \AT2 , \AIm) were calculated and 
 the results are shown in Fig.~\ref{fig:fsval}.
Significant bias is seen in the angular 
observable for an S-wave magnitude of greater than 5\%.
The linear increase in the bias 
is another consequence of the (1-\FS) factor.

\subsection{Measuring the S-wave in \BdToKpill}

Obtaining unbiased values for the angular observables beyond the limits
 shown requires a measurement of the S-wave contribution rather than ignoring it.
With the formalism developed in Sect.4, three options are explored for measuring
this. The first option is to ignore the \psq dependence and simply fit for
\psq-averaged values of \FS and \AS. The second option is to fit the \psq 
 line-shape simultaneously with the angular distribution.
This can be done in a small $p$ window between $800$ and $1000\mev$
or in the region from the lower kinematic threshold 
to $1200\mev$.
In all cases the datasets used to perform the 
studies are identical to those used in Sect. 5.1.
The difference is in how the fit is performed.
In each case, the dataset and the S-wave sizes 
refer to the number of events in the smaller \psq window.

The angular distribution without \psq dependence is given in Eq.~\ref{eq:theo4dint}.
for each set of samples, we look at the resolution, the mean 
and the width of the pull distribution of the angular observables. 

The change in the resolution obtained on the angular observables 
for the three methods of including the S-wave in the angular distribution
is demonstrated by plotting the ratio with respect
to the resolution obtained when a single 
P-wave state is assumed.

\begin{figure}[tb]
\centering
\subfigure[\AFB]{\includegraphics[width=0.48\textwidth]{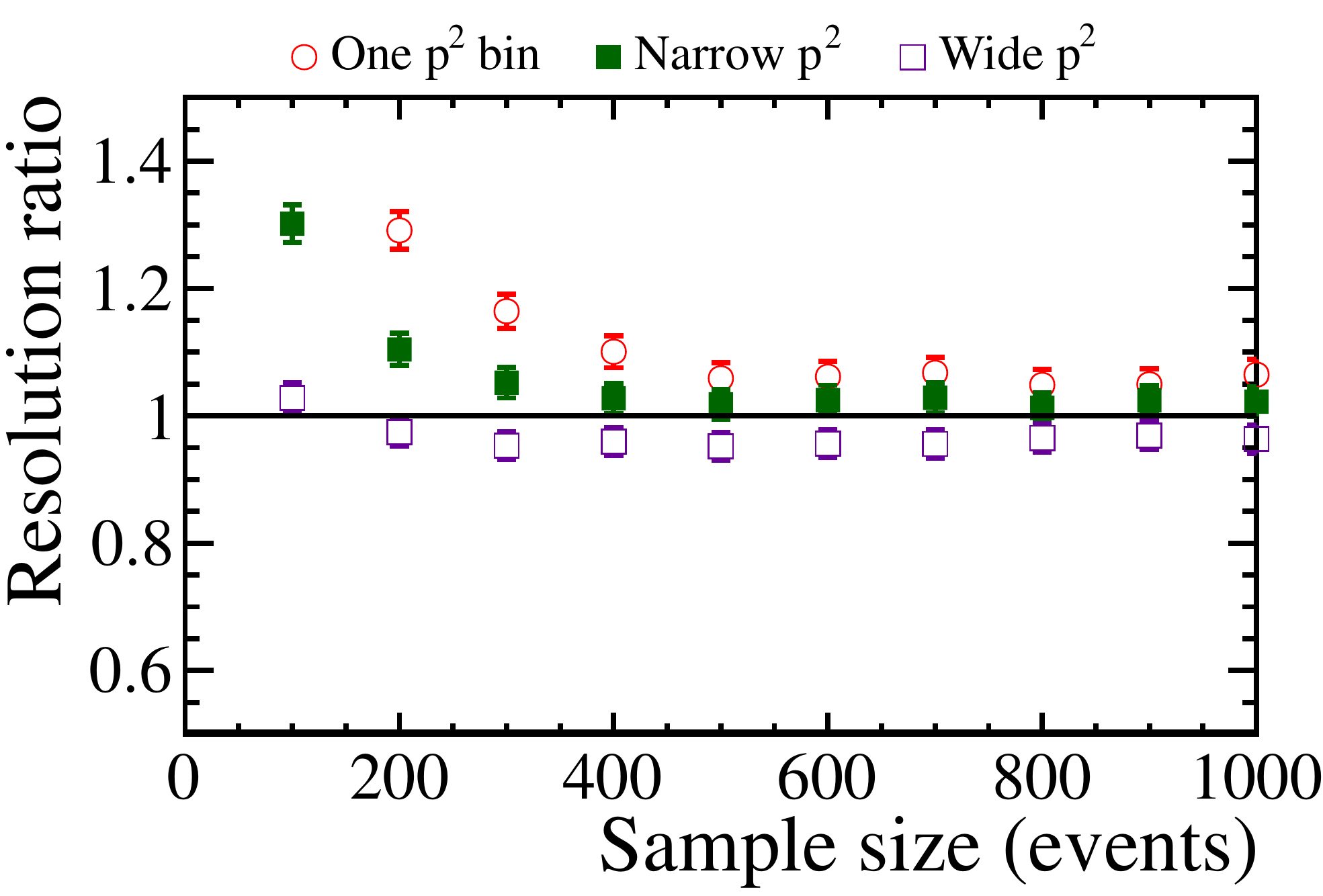}}
\subfigure[\FL]{\includegraphics[width=0.48\textwidth]{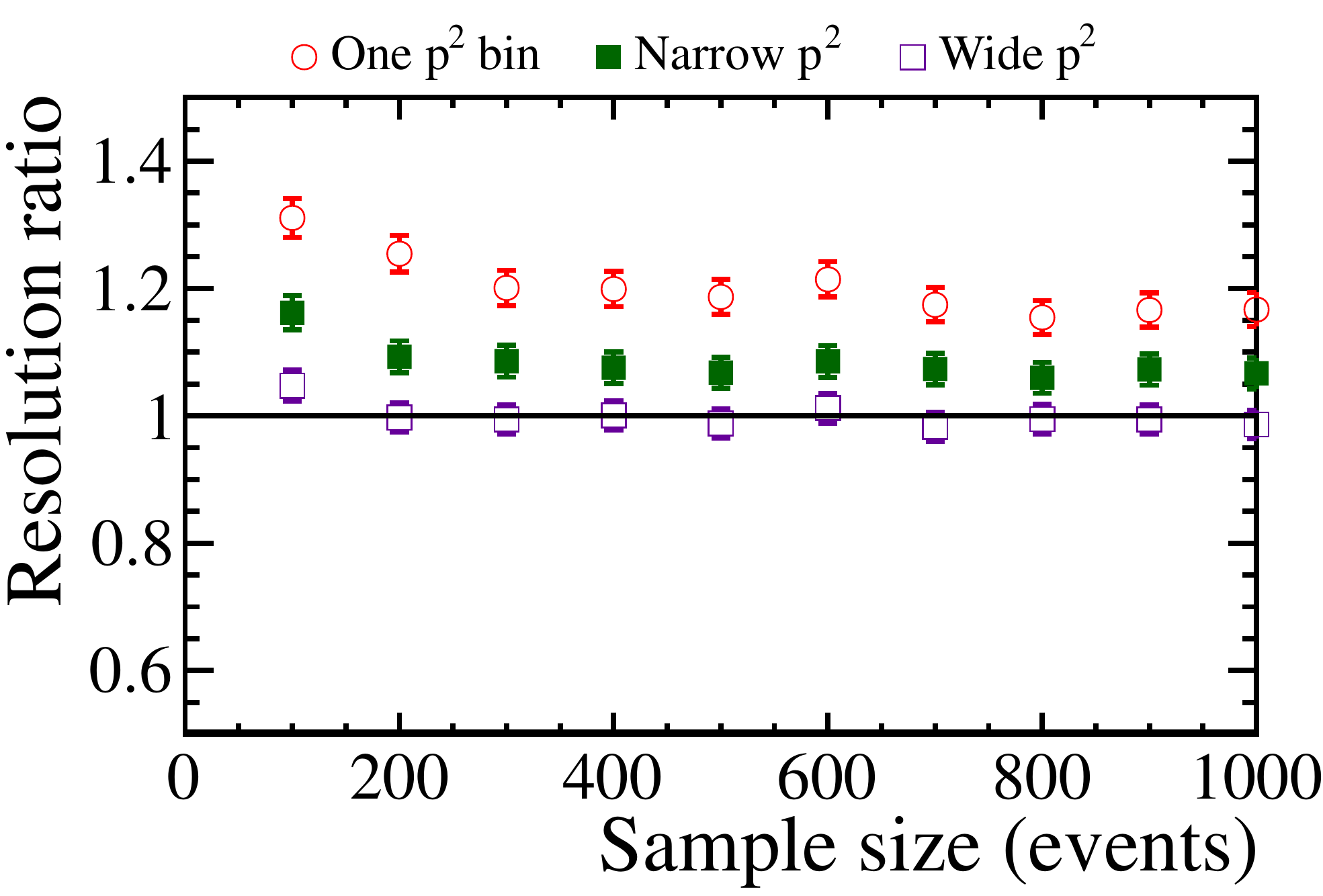}}
\subfigure[\AT2 ]{\includegraphics[width=0.48\textwidth]{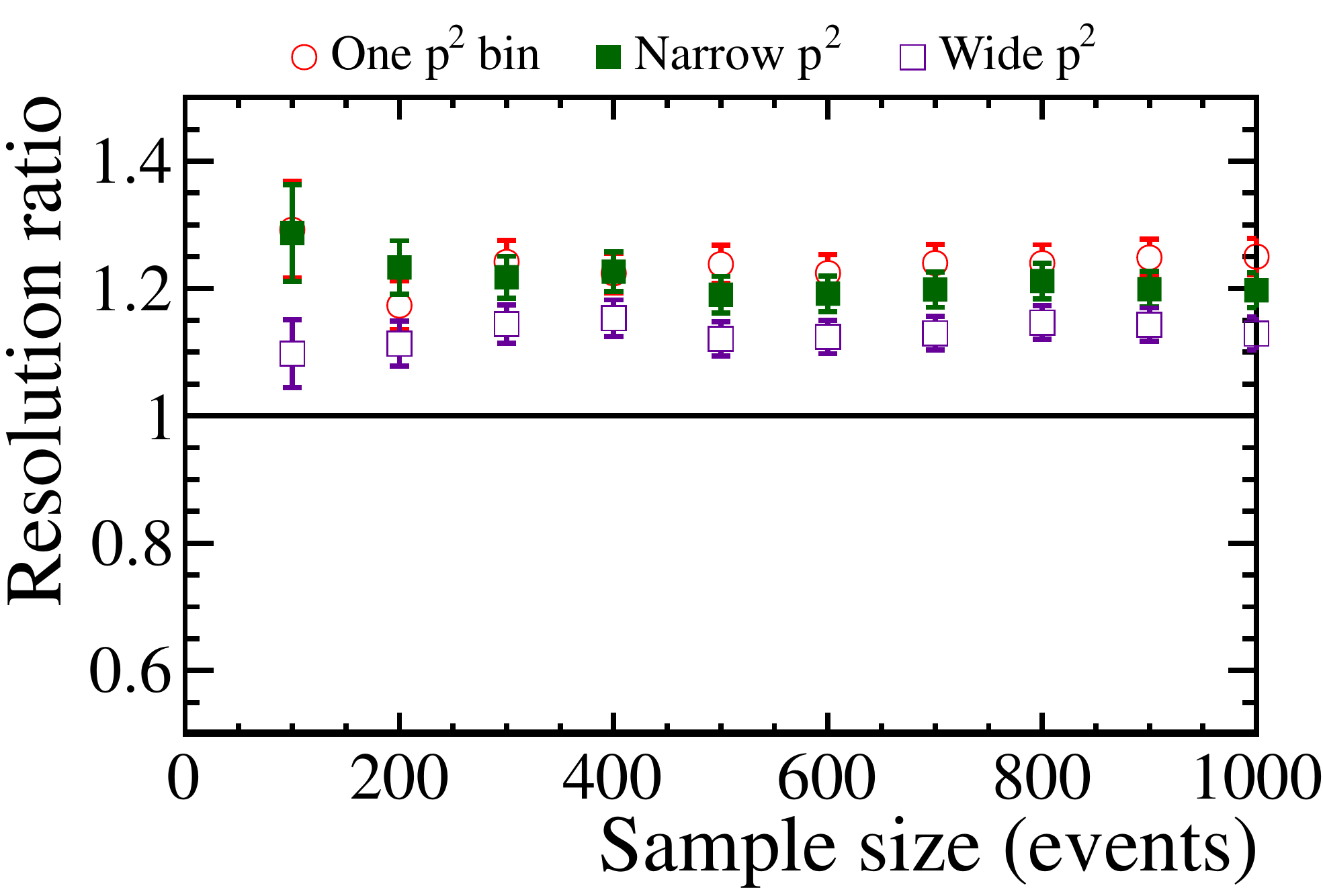}}
\subfigure[\AIm]{\includegraphics[width=0.48\textwidth]{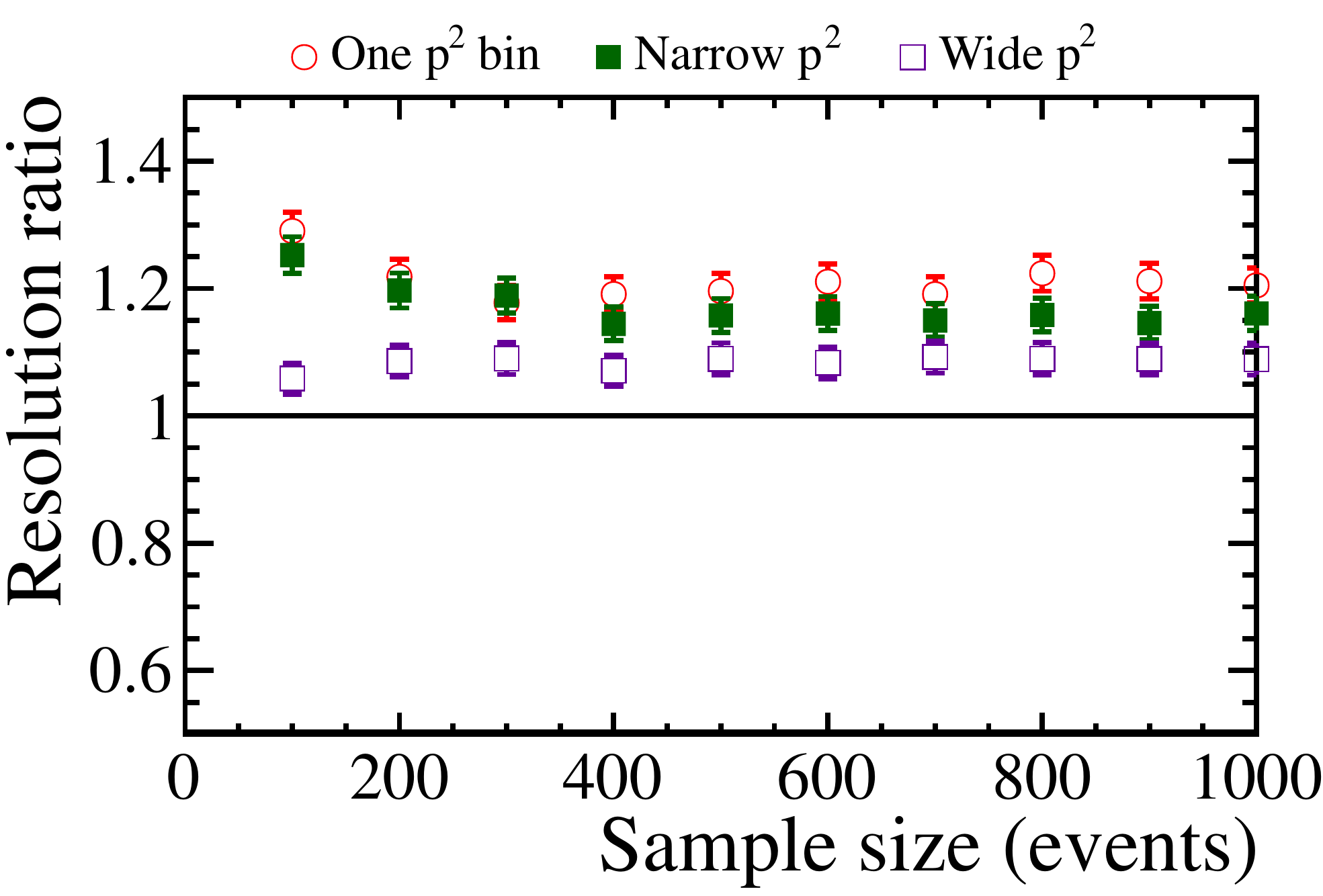}}
\caption{Resolutions for three different methods to incorporate 
the S-wave relative to the resolution obtained when the S-wave is ignored. 
It can be seen that the best resolution is obtained when using the largest \psq window.
The original resolution is recovered to within 10\%. ~\label{fig:ratiods}}
\end{figure}

\begin{figure}[tb]
\centering
\subfigure[\AFB]{\includegraphics[width=0.48\textwidth]{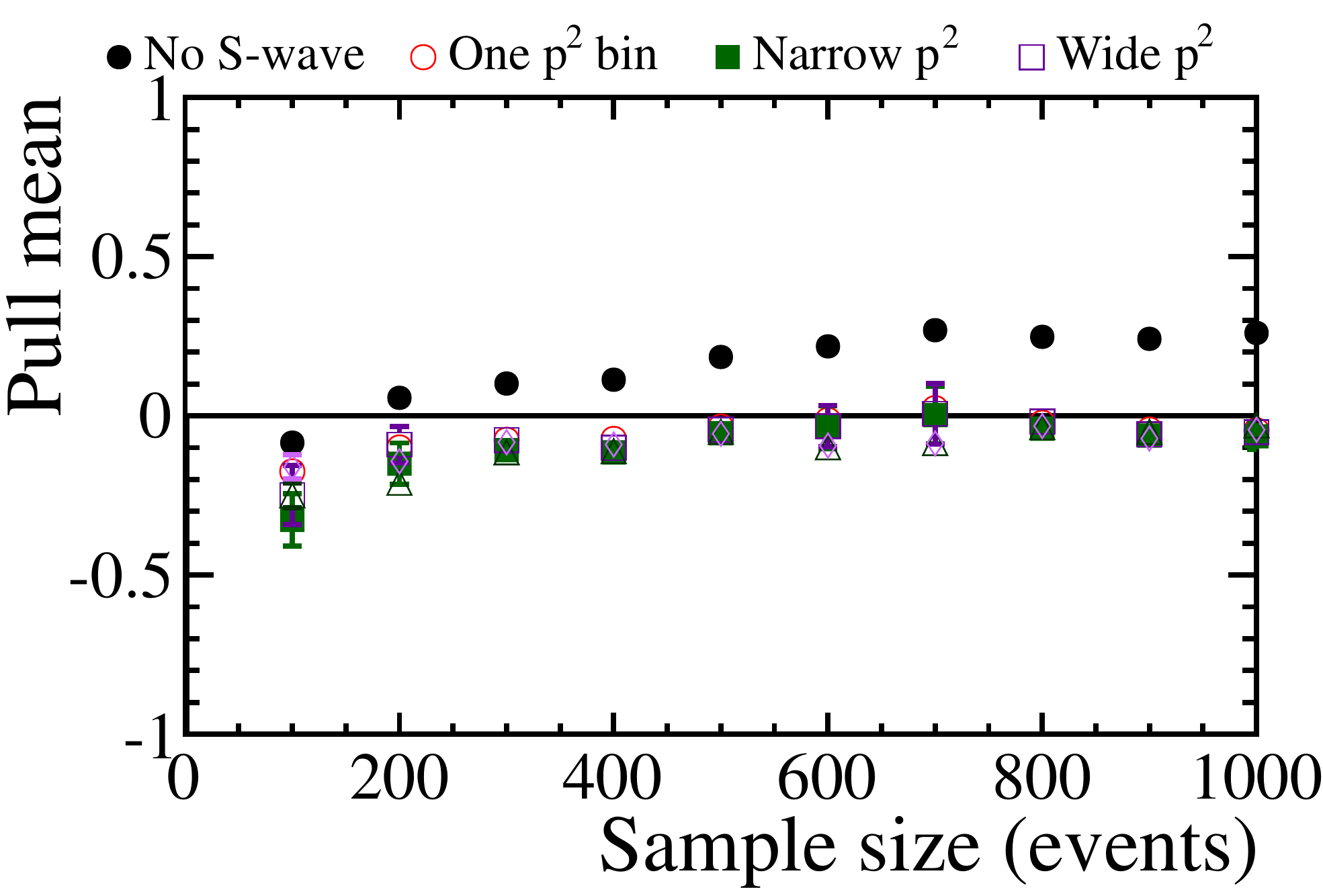}}
\subfigure[\FL]{\includegraphics[width=0.48\textwidth]{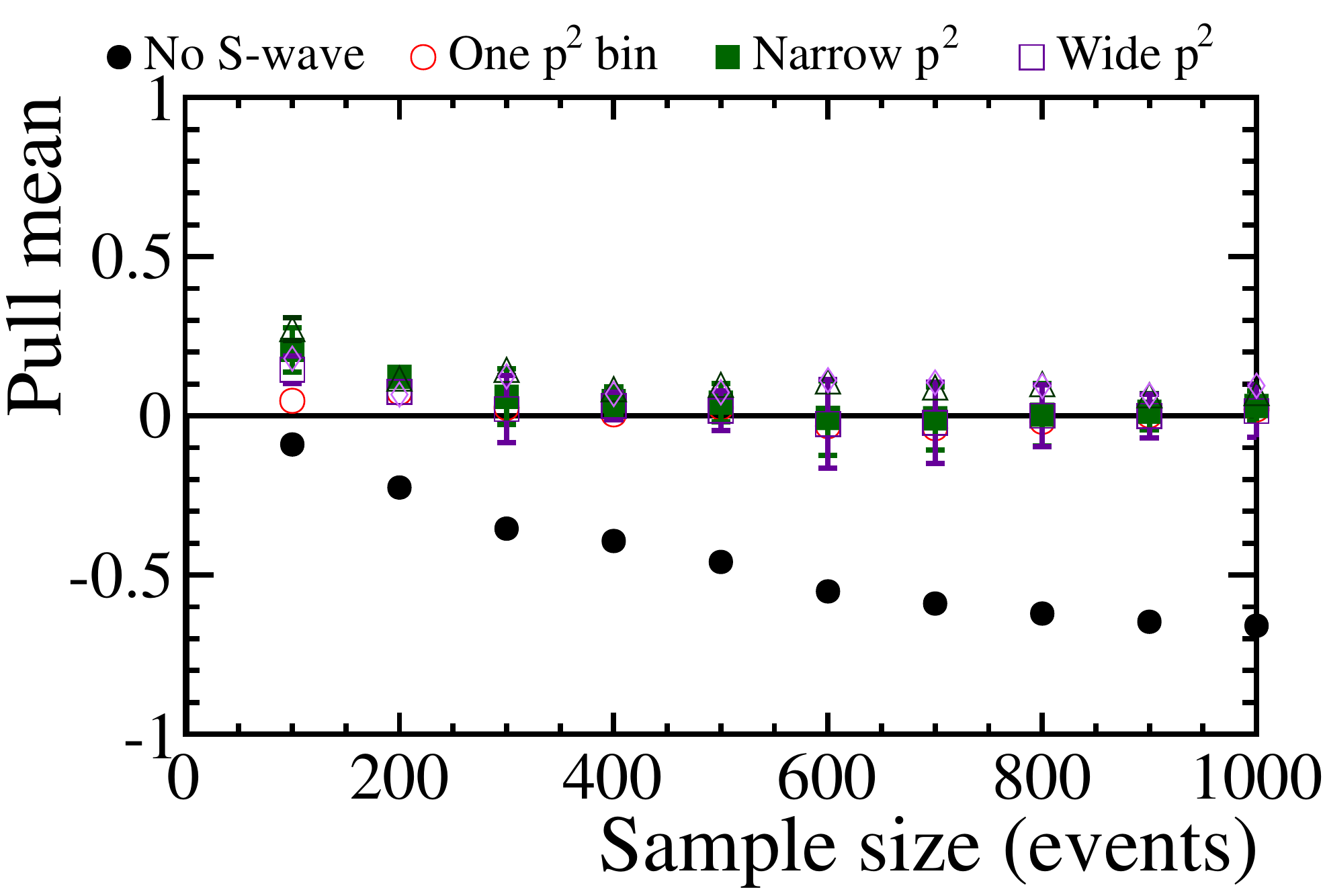}}
\subfigure[\AT2 ]{\includegraphics[width=0.48\textwidth]{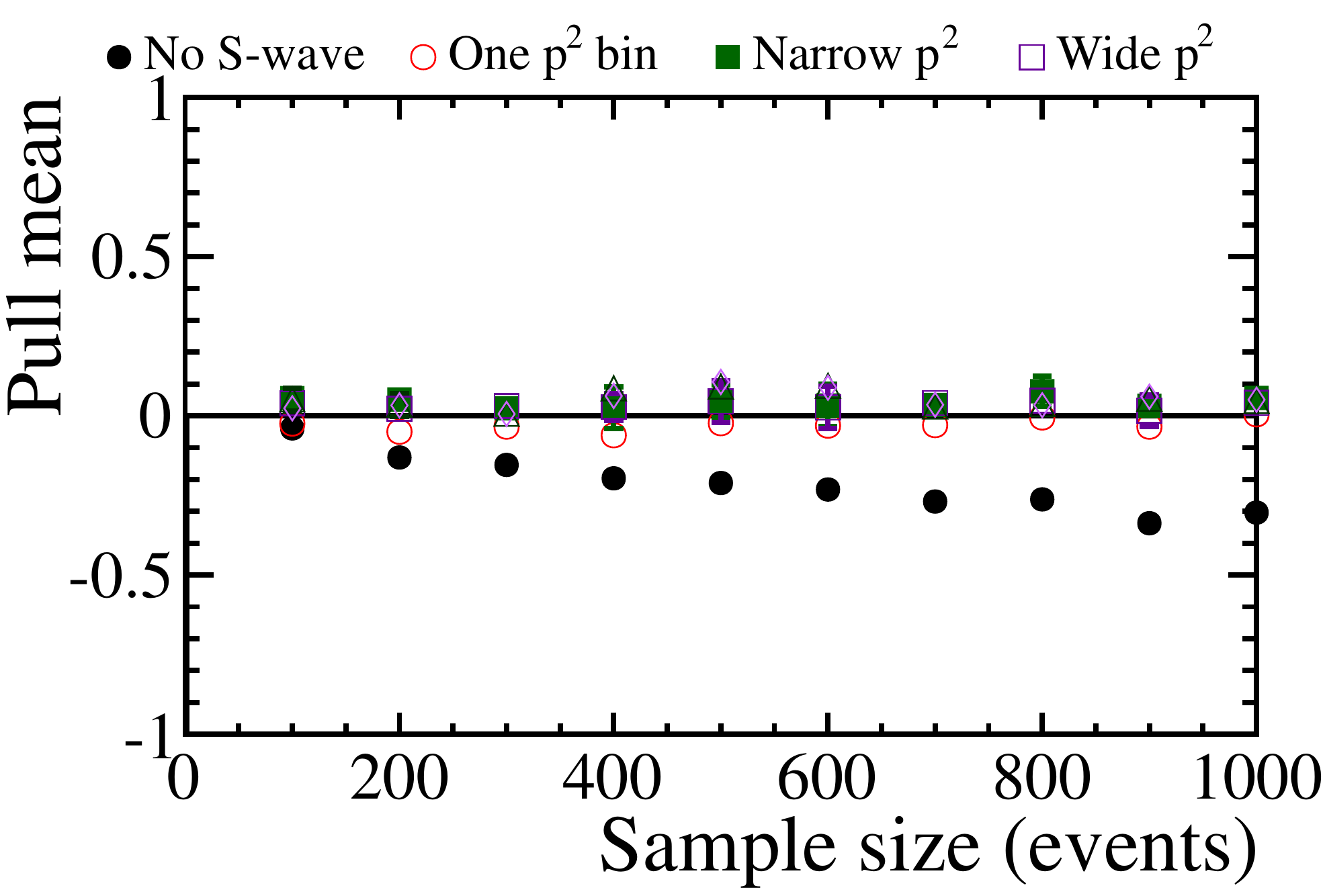}}
\subfigure[\AIm]{\includegraphics[width=0.48\textwidth]{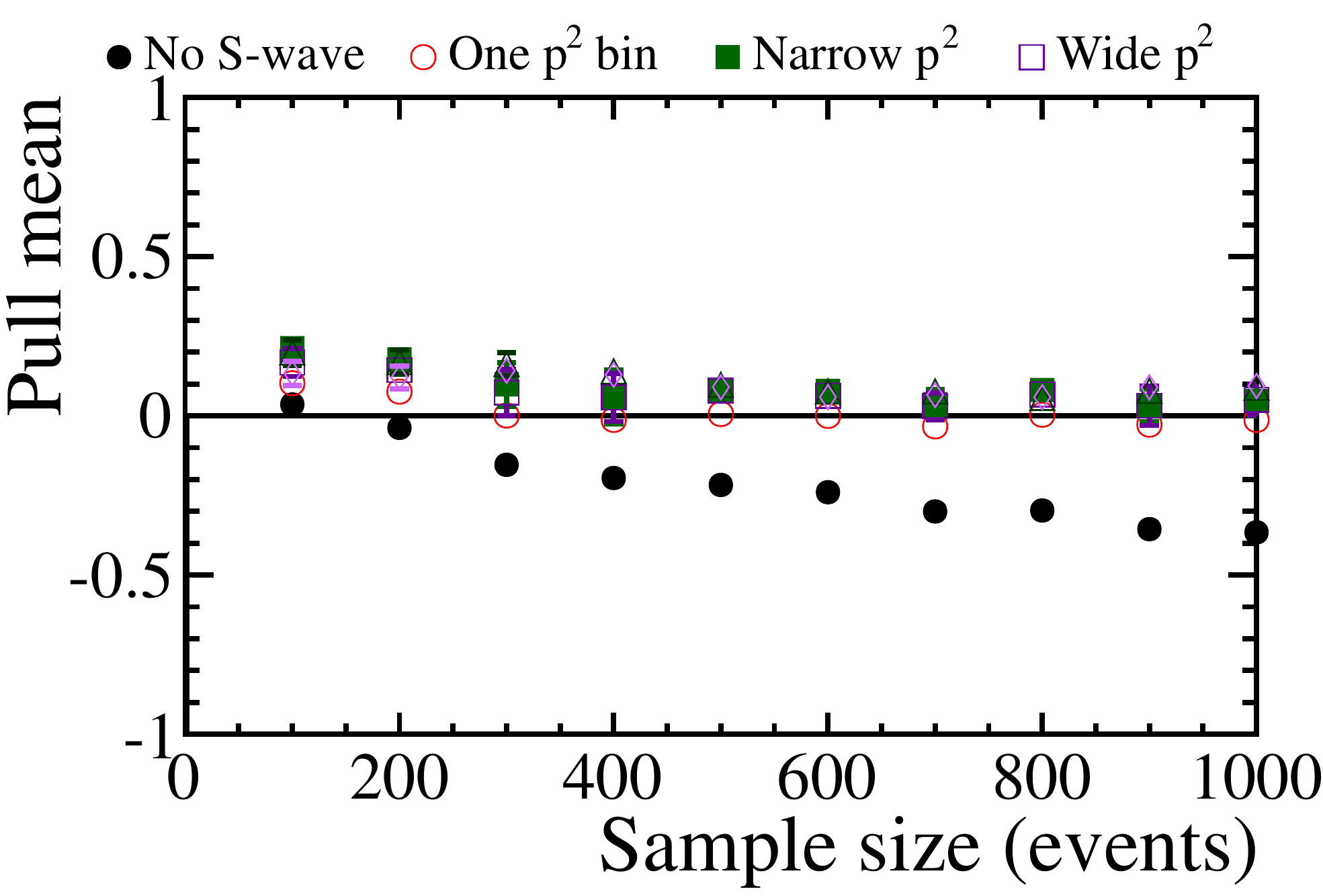}}
\caption{Pull mean for the three different  
methods to incorporate the S-wave and when the S-wave 
is ignored. There is a slight bias when the S-wave is 
included for datasets of less than 200 events but this 
bias is removed from  all the observables 
when the S-wave is included in the fit 
for datasets of over 500 events. ~\label{fig:combods}}
\end{figure}

The resolutions and the mean of the pull distributions for the three different fit methods 
(ignoring the \psq dependence, fitting a narrow \psq window and fitting a wide \psq window)
relative to the resolution and mean obtained using the assumption of a pure P-wave state. 
The ratio between the fit methods including the S-wave in angular distribution and 
assuming a P-wave state as a function of dataset size
are shown in  Fig~\ref{fig:ratiods}.
The pull mean for all four fit methods is shown in Fig~\ref{fig:combods}.

For all observables, it can be seen that the resolution degrades when
the S-wave is included and the \psq dependence is ignored.  The
resolution degrades by a smaller amount when the \psq dependence is
included in a small bin and the original resolution is recovered to
within 10\% when using the large \psq range.  There are two effects
contributing to the improvement of the resolution. There are more
P-wave events in the larger range and the wider mass window allows for
the S-wave to be constrained by using the information from above and
below the P-wave resonance.  This results in the best resolution when
the S-wave is included in the angular distribution.

For all the observables, the pull mean approaches zero for datasets of
greater than 300 events implying that the bias present in all the
observables when a pure P-wave state is removed when an S-wave is
included in the angular distribution. This means that the inclusion of
the S-wave component will be mandatory for all future experimental
analyses.

Another approach to reduce the bias from the S-wave is to ignore it in
fits but to only include data from a narrower window in $p$ arounnd
the $\Kstarz(892)$ resonance. By reducing the window from 200\mev to
100\mev, the P-wave component is reduced by 20\% while the S-wave
component is roughly halved. Conducting the same tests as described
above shows, as expected, a 10\% increase in the statistical error of
the observables while the bias for a given dataset is reduced by a
factor two. Given what has been shown in this paper, the experimental
datasets will in the future be so large that the best approach is to
fit the S-wave rather than half the bias and accepting an increased
statistical uncertainty.

Until now the lineshape of the S-wave has been parameterised according
to the LASS model (Eqs.~\ref{eq:LASS1}-\ref{eq:LASS3}). We asses the
model dependence of this assumption by using the alternative isobar
model~\cite{PhysRevLett.89.121801} for generating the S-wave component
while keeping the same fit model. This only has an effect on the fits
where a fit is performed to the $p^2$ dependence. The results of this
show that the systematic uncertainly due to the model dependence is
much smaller than the statistical error for all observables for all
sample sizes we studied.

\section{Conclusion}

In summary, the inclusion of a resonant \kpi S-wave in the angular
analysis of \BdToKstll has been formalised and the complete angular
distribution for both an S- and P-wave state described.  We find that
the inclusion of an S-wave state has an overall dilution effect on the
theoretical observables.  The impact of an S-wave on an angular
analysis is evaluated using toy Monte Carlo datasets.  We find that
the S-wave contribution can only be ignored for datasets of less than
200 events.  The bias on the angular observables incurred by assuming
a pure P-wave \kpi state can be removed by including the S-wave in the
angular distribution.  The degradation in resolution on the angular
observables from fitting a more complicated angular distribution can
be minimised by performing the fit in a wide region around the
$\Kstarz(892)$ resonance. The systematic uncertainty introduced by the
model dependence of the S-wave lineshape is minimal and can be
ignored.

\section*{Acknowledgements}

\noindent T.B. acknowledges support from \cern. 
U.E and A.S acknowledge support from the Science and Technologies Facilities Council 
under grant numbers ST/K001280/1 and ST/F007027/1.

\bibliographystyle{LHCb}
\bibliography{main}

\end{document}